\documentclass[lettersize,journal]{IEEEtran}
\usepackage{amsmath,amsfonts}
\usepackage{algorithmic}
\usepackage[caption=false,font=normalsize,labelfont=sf,textfont=sf]{subfig}
\usepackage{textcomp}
\usepackage{stfloats}
\usepackage{url}
\usepackage{verbatim}
\usepackage{graphicx}
\def\BibTeX{{\rm B\kern-.05em{\sc i\kern-.025em b}\kern-.08em
    T\kern-.1667em\lower.7ex\hbox{E}\kern-.125emX}}
\usepackage{balance}

\usepackage{booktabs}  
\usepackage{graphicx}
\usepackage[switch]{lineno}
\usepackage{amsmath}
\usepackage{amsthm}
\usepackage{amssymb}
\usepackage{multirow}
\usepackage{bbding}
\usepackage{algorithm}
\usepackage{algorithmic}
\usepackage{xspace}
\usepackage{arydshln}
\usepackage[inline]{enumitem}

\usepackage{subcaption}

\usepackage{colortbl}
\usepackage{xcolor}         

\definecolor{healthcolor}{HTML}{B3ECFF}

\newcommand{\ie}{{i.e.,}\xspace}

\newcommand{\eg}{{e.g.,}\xspace}
\newcommand{\baby}{{MitiHarm}\xspace}

\newcommand{\fig}{Fig.\xspace}

\begin{document}
\title{Measuring, Revealing and Mitigating Harmful Content Propagation in Recommender Systems}

\author{Xiaokun Zhang, Youlin Wu, Junyu Lu, Jiamin Chen, \\ Zhaochun Ren, Hongfei Lin, Fenglong Ma, Xue Liu,~\IEEEmembership{Fellow,~IEEE}, and Chen Ma
\IEEEcompsocitemizethanks{
\IEEEcompsocthanksitem This work is supported in part by the Early Career Scheme (No.CityU 21219323) and the General Research Fund (No.CityU 11220324) of the University Grants Committee (UGC), the NSFC Young Scientists Fund (No.9240127), the Donation for Research Projects (No.9229164 and No.9229216), and the Key Laboratory of Social Computing and Cognitive Intelligence (Dalian University of Technology), Ministry of Education (SCCI2025YB04). (Corresponding author: Chen Ma.)
\IEEEcompsocthanksitem Xiaokun Zhang,  Jiamin Chen, Chen Ma are with the Department of Computer Science, City University of Hong Kong, Hong Kong, China. Xiaokun Zhang is also supported by the Key Laboratory of Social Computing and Cognitive Intelligence (Dalian University of Technology), Ministry of Education.
E-mail: dawnkun1993@gmail.com, jmchen26-c@my.cityu.edu.hk, chenma@cityu.edu.hk
\IEEEcompsocthanksitem Youlin Wu, Junyu Lu, Hongfei Lin are with the School of Computer Science and Technology, Dalian University of Technology, Dalian, China.
E-mail: \{wuyoulin,dutljy\}@mail.dlut.edu.cn, hflin@dlut.edu.cn
\IEEEcompsocthanksitem Zhaochun Ren is with the Leiden Institute of Advanced Computer Science, Leiden University, Leiden, Netherlands. E-mail: z.ren@liacs.leidenuniv.nl
\IEEEcompsocthanksitem Fenglong Ma is with the College of Information Sciences and Technology, Pennsylvania State University, Pennsylvania, USA. E-mail: fenglong@psu.edu
\IEEEcompsocthanksitem Xue Liu is with the Department of Computer Science, MBZUAI, Abu Dhabi, UAE, and with the School of Computer Science, McGill University, Montreal, Canada. E-mail: steve.liu@mbzuai.ac.aeg
}

}


\maketitle

\begin{abstract}
Recommender systems help users navigate vast amounts of information by filtering content according to their historical behaviors. As these systems increasingly shape everyday information access, they largely determine the content users are exposed to. However, current research typically prioritizes catering to user interest, while overlooking the existence and propagation of harmful content, such as violence, hate speech, and pornography. In this work, we take a systematic step toward investigating harmful content in recommender systems from three aspects: measuring, revealing, and mitigating. First, we construct datasets with harmfulness-related annotations and present new evaluation metrics to quantitatively measure the harmfulness of recommendation results. Second, our empirical analysis reveals that users tend to engage with harmful content and prevailing recommendation methods often contribute to the proliferation of such content. Third, to address this issue, we introduce a generic framework, \baby, which integrates harmfulness-related signals into both item and user representation learning. Experiments show that \baby can significantly mitigate the propagation of harmful content while preserving user satisfaction. Resources of this work are available at \url{https://github.com/Zhang-xiaokun/MitiHarm}.
\end{abstract}

\begin{IEEEkeywords}
Responsible AI, Harmful Content Propagation, Recommender Systems, Benchmarks, Empirical Findings, Mitigation Framework.
\end{IEEEkeywords}

\section{Introduction}
\IEEEPARstart{I}{n} the digital age, people are exposed to an unprecedented volume of information, far exceeding their capacity for timely consumption, which gives rise to the information overload~\cite{Dang@TPAMI2026, Dong@TPAMI2025}. For instance, users are routinely confronted with thousands of news articles, endless social media feeds, and a constant stream of short videos across diverse media platforms. To address this issue, recommender systems (RS) have been developed to selectively filter relevant information for users by analyzing their historical behaviors~\cite{IKNN, Abhinandan@WWW2007}. Owing to their effectiveness in meeting user needs, RS have become ubiquitous across a wide range of domains, particularly in online content services such as news platforms~\cite{Wu@RecSys2025}, digital reading services~\cite{Gupta@ICML2021}, and short video applications~\cite{Huang@WWW2024}. 

As RS continue to shape user experiences, their role has evolved far beyond simple content filtering. In many scenarios, they serve as the primary gateway for information access and propagation. For instance, more than 30\% of browsing behaviors on Amazon\footnote{\url{https://www.amazon.com/}} shopping site are triggered by RS~\cite{SharmaHW15}, nearly 80\% of movie consumption on Netflix\footnote{\url{https://www.netflix.com/}} website originates from recommendations~\cite{HallinanS16}, and over 40\% of news reading on Google News\footnote{\url{https://news.google.com/}} platform is guided by recommendation algorithms~\cite{DasDGR07}. Even more strikingly, on popular short-video applications, such as Youtube\footnote{\url{https://www.youtube.com/}} and Douyin\footnote{\url{https://www.douyin.com/}}, users typically consume content simply by swiping the screen, with each successive video automatically selected and delivered by the underlying algorithm. Together, these facts illustrate that recommender systems directly determine much of the content users are exposed to and engage with on a daily basis.

\begin{figure*}[t]
  \centering
  \includegraphics[width=0.99\linewidth]{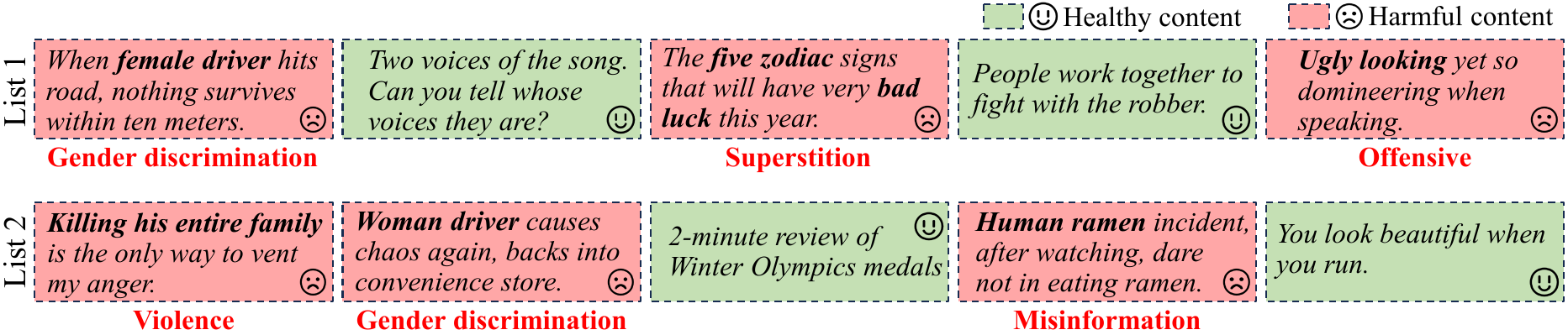}
  \caption{Recommendation lists generated by a representative model, BERT4Rec, where harmful content is prevalent.}\label{intro_case}
\end{figure*}

However, current recommendation methods are dedicated to improving prediction accuracy, striving to align recommendations with user interest inferred from their historical behaviors~\cite{Kong@NIPS2024, Zhang@TKDE2025, BiPNet}. While this accuracy-oriented manner has advanced the field significantly, it overlooks the harmfulness of the recommended content. To illustrate this issue, we randomly sample 100 recommendation lists generated by a widely used method, BERT4Rec~\cite{BERT4Rec}, and observe that \textit{nearly half of the recommended items contain harmful content}, such as gender discrimination, misinformation, and offensive language, as shown in~\fig~\ref{intro_case}. Exposure to such content, especially at scale, can negatively affect users’ mental well-being, with heightened risks for vulnerable groups like children and adolescents~\cite{Arnav@CSUR2024,ulbricht2024harmful}. Moreover, the widespread circulation of harmful content could pose serious risks to society, including fueling misinformation, exacerbating social divisions, and eroding public trust in digital ecosystems~\cite{Elahi@AIE2022, Miriam@WEBSCI2024}. This concerning reality raises a critical question:\textit{ Do recommender systems facilitate the propagation of harmful content, and if so, how can such propagation be effectively mitigated?}

Unfortunately, few existing studies have systematically examined recommender systems from the perspective of content harmfulness. A major obstacle lies in the absence of standardized benchmarks. On one hand, existing datasets lack annotations relevant to content harmfulness, like whether a piece of content is healthy or harmful. On the other hand, there is no established metric to quantitatively evaluate the harmfulness of recommendation results, making it difficult to assess or compare methods in this regard. This absence of essential resources significantly hinders progress toward developing responsible recommender systems.

To this end, this work aims to investigate this critical yet under-explored dimension of RS through the following three key aspects. (\romannumeral1) \textbf{Benchmark Development}. We begin by clarifying the definition of harmful content based on the consensus of sociology and psychology~\cite{Arnav@CSUR2024,ulbricht2024harmful}. Building on this, we enrich several content recommendation datasets with harmfulness-related annotations, including binary harmfulness tags (\ie healthy vs. harmful) and corresponding reasons that provide justifications for the assigned tags. To support this effort, we devise a LLM-human collaborative annotation paradigm: large language models (LLMs) first generate preliminary harmfulness tags and reasons, which are then reviewed and revised by human annotators.
Moreover, we present two new metrics, Healthy content proportion (HP) and Healthy content ordering (HO), to quantitatively measure the harmfulness of recommendation results from different aspects. (\romannumeral2) \textbf{Empirical Findings.} Using the constructed benchmarks, we reveal the widespread presence of harmful content in real-world data. We also find that users tend to engage with harmful content. In addition, our evaluation of representative recommendation models shows that existing methods consistently facilitate the propagation of harmful content. These findings expose a critical safety-related shortcoming in the design of contemporary recommendation systems. (\romannumeral3) \textbf{Mitigation Framework.} To mitigate this issue, we introduce \baby, a generic and model-agnostic framework that integrates harmfulness-related signals into both item and user representation learning, enabling RS to generate recommendations that prioritize healthy content while aligning with user preferences.
In summary, this work makes the following key contributions:
\begin{itemize}[leftmargin=*]
    \item \textbf{Benchmark Development}. We propose a novel benchmark for evaluating recommendation harmfulness. Three datasets annotated with fine-grained harmfulness-related information are constructed, and two new metrics, HP and HO, are designed to quantitatively examine recommender systems in terms of recommendation harmfulness.
    \item \textbf{Risk Identification}. We reveal the overlooked risk of existing recommender systems in spreading harmful content. To the best of our knowledge, this study marks the first systematic effort to investigate recommender systems from the harmfulness perspective.
    \item \textbf{Framework Proposal}. We present \baby, a generic, model-agnostic framework that effectively mitigates the propagation of harmful content in recommender systems while preserving user satisfaction.
\end{itemize}

\section{Benchmark Development}\label{sec:bechmark}

\subsection{Dataset Construction}

\subsubsection{Definition of Harmful Content}\label{harm_definition}
We first clarify the definition of harmful content as follows: \textit{\textbf{Harmful content} refers to the text that causes or promotes harm to individuals or groups, or involves or encourages violations of laws or widely accepted social norms. This includes, but is not limited to, content that involves or encourages violence, hate speech, misinformation, offensive language, pornography, or other forms of socially inappropriate or harmful material}. This definition is grounded in five key considerations: (a) common practices in existing harmful content literature~\cite{Arnav@CSUR2024}; (b) perspectives widely accepted in sociology and psychology~\cite{ulbricht2024harmful}; (c) content policies and community guidelines specified in the Terms and Conditions of major social media platforms, like Sina WeiBo\footnote{\url{https://m.weibo.cn/}} and Twitter\footnote{\url{https://x.com/}}; (d) prompt efficiency and clarity for LLM-based annotation; and (e) our preliminary analysis of existing datasets. More discussion about harmful content definition is presented in Appendix~\ref{app_definition}.

\subsubsection{Harmfulness-related Labeling}\label{sec:health_label}
\begin{figure*}[t]
  \centering
  \includegraphics[width=0.98\linewidth]{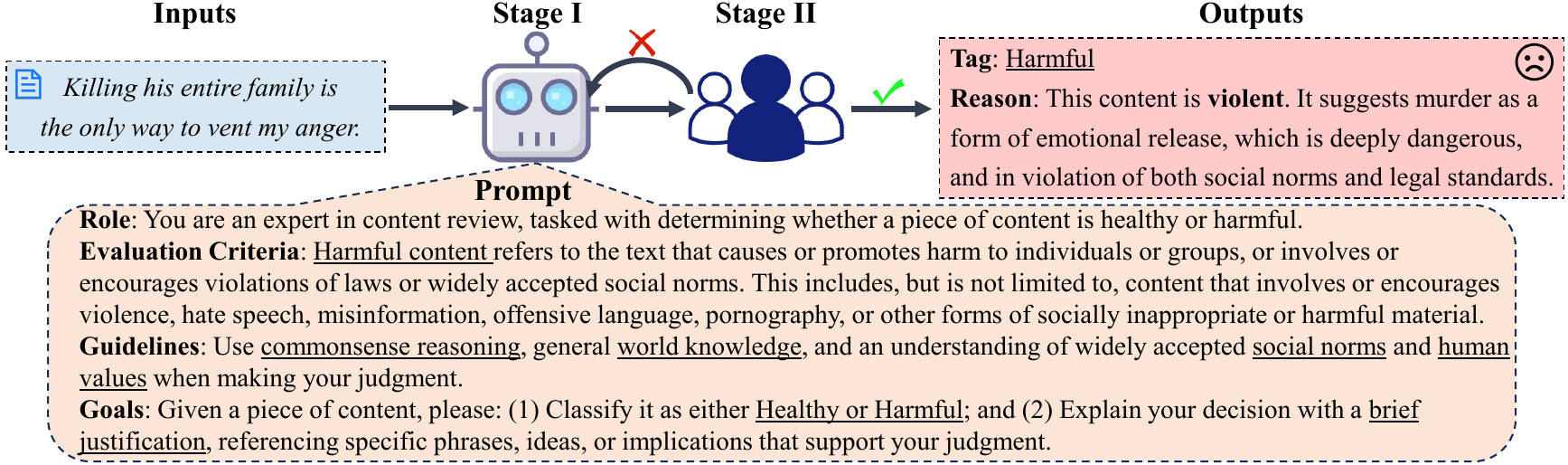}
  \caption{Illustration of LLM-human collaborative annotation paradigm which consists of two sequential stages. Stage I: LLM Annotation Stage, and Stage II: Human Revision Stage.}\label{annotation}
\end{figure*}

In this work, we mainly annotate the following two types of harmfulness-related information: (a) \textbf{Harmfulness tag}. A binary label that classifies content as either healthy or harmful, serving as a fundamental indicator of its harmfulness; and (b) \textbf{Harmfulness reason}. A textual explanation that provides the detailed reason supporting the harmfulness classification. It describes characteristics about content harmfulness beyond the binary label. Specifically, harmful themes are often expressed through subtle or implicit cues, including metaphors, sarcasm, or exaggeration, making them difficult to detect~\cite{Binny@AAAI2021, Mou@CIKM2020}. The harmfulness reason explicitly articulates these clues, serving as a clear and informative signal of content harmfulness. An example of the annotation is presented in~\fig~\ref{annotation}.

\subsubsection{Basic Datasets}
This work extends three public content recommendation datasets, \textbf{TN}, \textbf{QB}~\cite{NineRec}, and \textbf{EB-NeRD}~\cite{NeRD}, to conduct the investigation on harmfulness of recommender systems. These datasets comprise diverse online textual content, including news articles and advertisements, and record user interactions with this content. For each dataset, an item's content is derived by combining its textual features, such as title, abstract, and full content. Details of these datasets are provided in Appendix~\ref{app_data}.

\subsubsection{Data Annotation}\label{section:data_anotation}
Large Language Models (LLMs), pre-trained on vast text corpora, possess rich world knowledge, demonstrate excellence in content understanding, and often align well with human values~\cite{Kong@NIPS2024,FineRec,Ouyang@NIPS2022}. Recent studies~\cite{Xiao@EMNLP2023,Zhang@EMNLP2023, Wang@CoRR2024} have also shown their great potential in supporting data annotation tasks. Inspired by this, we devise a LLM-human collaborative annotation paradigm to conduct the harmfulness annotation. This approach aims to enhance both annotation quality and efficiency simultaneously. As illustrated in~\fig~\ref{annotation}, it consists of the following two sequential stages:

\textbf{LLM annotation stage.} We first employ LLMs to generate annotations for both the harmfulness tag and corresponding reason for each item. To guide the annotation process, we crafted a prompt tailored for harmfulness evaluation. As shown in~\fig~\ref{annotation}, this prompt specifies several instructional elements, including Role definition, Evaluation Criteria, Guidelines, and Goals, which collectively guide the LLM to attend specifically to content harmfulness assessment. Noted that the Evaluation Criteria are tightly aligned with our definition of harmful content, ensuring consistency with widely accepted standards. Furthermore, we explicitly require LLMs to output both the binary harmfulness tag and a supporting textual reason in a single pass. This operation encourages the LLM to think about its harmfulness assessment deeply, similar to the chain-of-thought prompting paradigm~\cite{Wei@NIPS2022}, contributing to improving LLMs' reasoning ability. More details about the implementation of LLM annotation stage are discussed in Appendix~\ref{app_LLManotation}.

\textbf{Human revision stage.} We invite five human annotators to review and revise the harmfulness annotations generated by LLMs. Each annotator independently evaluates the correctness of the generated tag. If the majority votes in favor of the LLM's annotation, the tag is accepted. Otherwise, we revise the tag according to the assessment of most annotators. For the harmfulness reason, if most annotators agree with both the tag and its explanations, the output is accepted. If not, the content is resubmitted to the LLM with the corrected tag to generate a reason using the prompt: ``\textit{The following content [content] is [corrected tag]. Please provide the reason for this assessment.}'' In practice, around 90\%, 85\%, and 79\% of the LLM-generated outputs are accepted without revision in TN, QB, and EB-NeRD, respectively, indicating the strong capability of LLMs in this annotation task. 

In summary, the devised LLM-human collaborative annotation paradigm proves to be both effective and efficient, demonstrating its strong potential for extending existing recommendation datasets with harmfulness-related annotations while substantially reducing human workload. 
Moreover, we provide several illustrative examples of the annotated results across three datasets in Appendix~\ref{app_IO}. Further details about the annotation process are presented in Appendix~\ref{app_ac} and ~\ref{app_error}.

\subsubsection{Annotator Selection and Training}\label{app_annotator}

\begin{table}[t]
\tabcolsep 0.2in
\center
\caption{Annotators demographics.}
\begin{tabular}{cc} 

\toprule 
\textbf{Characteristic} & \textbf{Demographics}\\  
\midrule 
Gender & 3 Female, 2 Male    \\ 
Age & $<$25: 2, $\geq$ 25: 3 \\  
Race \& Region & 3 Asian, 2 others  \\
Education & 3 M.D., 2 Ph.D.  \\
\bottomrule
\end{tabular}
\label{anno_tors}%
\end{table}

To ensure a diverse and inclusive perspective in the human revision process, annotators are selected based on varied demographics, including gender, age, ethnicity, geographical region, and education level. The detailed demographics of the annotators are provided in Table~\ref{anno_tors}. To ensure high-quality annotations, all annotators underwent a training process. Specifically, they completed three test rounds, each involving 100 items. In each round, annotators independently evaluated the outputs of LLM, while researchers finalized the annotation results. After each round, discussions were held to review errors and ambiguities. These sessions helped clarify annotation standards and led to the establishment of detailed guidelines, including how to handle edge cases. In addition, all annotators are required to be proficient in both English and Chinese, as the annotation process involves content in both languages.

\subsubsection{Data Statistics}

\begin{figure*}[t]
  \centering
  \includegraphics[width=0.9\linewidth]{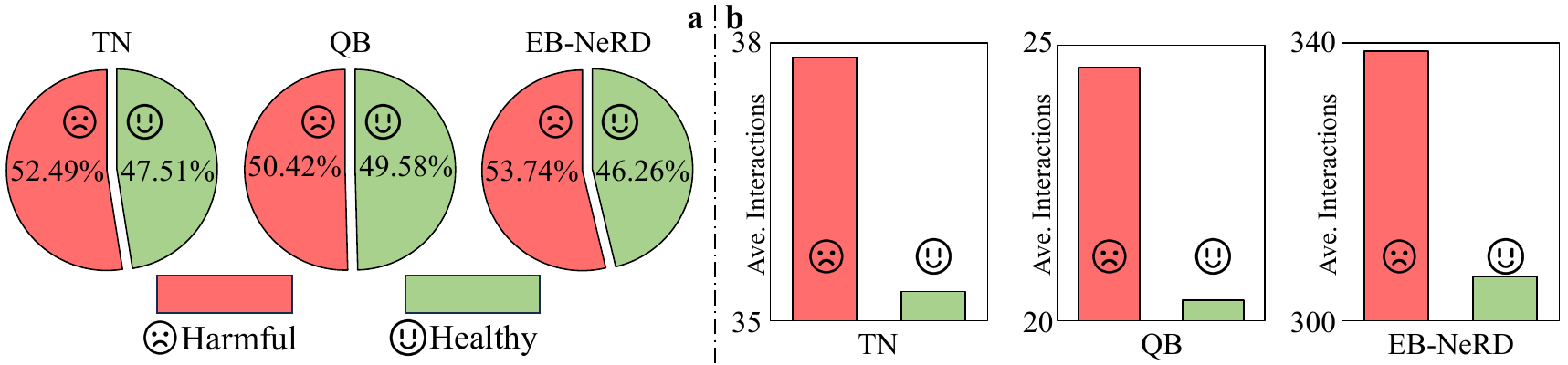}
  \caption{Empirical findings. (a) Distribution of harmful and healthy items across three datasets; and (b) Average number of user interactions per item for harmful and healthy content.}\label{findings}
\end{figure*}

\begin{table}
\center
\tabcolsep 0.15in
\caption{Statistics of datasets.}
\begin{tabular}{ccccc}
\toprule
Datasets      & TN & QB & EB-NeRD  \\
\midrule
\#user     & 20,211  & 17,689  & 18,827    \\
\#item        & 3,254   & 5,938   & 10,328    \\
\#interaction & 122,490  & 133,263 & 2,391,329  \\
avg.length    & 6.06     & 7.53     & 127.02    \\
\bottomrule
\end{tabular}
\label{statistics}%
\end{table}
For model training and evaluation, we adopt a standard next-item prediction protocol, where the last interacted item in each user behavior sequence is treated as the target (\ie ground truth), and the preceding interactions are used to model user behavior patterns~\cite{GRU4Rec,MMSBR,DIMO}. Following common practices~\cite{SR-GNN,Zhang@WSDM2026, DIDN, Zhang@PR2026}, we remove sequences containing only one single interaction and filter out items that appear fewer than five times. Each enriched dataset is then chronologically split into training, validation, and test sets with a ratio of 7:2:1. The statistical details of all datasets are summarized in Table~\ref{statistics}.

\subsection{Harmfulness Metric Formulation}\label{sec:healthmetric}

Existing recommendation metrics primarily evaluate whether items of user interest, \ie target items, are included in the recommendation list, either by measuring their coverage (\eg Precision) or considering their ranking positions (\eg MRR and NDCG). Inspired by this, we introduce the following two novel harmfulness metrics, 

\textbf{HP@$k$}, \underline{H}ealthy content \underline{P}roportion, measures the proportion of healthy content among the top-$k$ recommendations, offering a direct assessment of healthy content coverage.

\textbf{HO@$k$}, \underline{H}ealthy content \underline{O}rdering, further incorporates ranks of healthy content by rewarding models that prioritize healthy content higher in the recommendation list. 

Formally, given a test set $T$, these two harmfulness evaluation metrics are computed as follows,
\begin{align}
    \text{HP}@k &= \frac{1}{|T|}\sum_{r \in T}\frac{\sum_{i=1}^{k}H[r^i]}{k}, \\
    \text{HO}@k &= \frac{1}{|T|}\sum_{r \in T}(\sum_{i=1}^{k}\frac{H[r^i]}{i}/\sum_{i=1}^{k}\frac{1}{i}),
\end{align}
where $r$ denotes the top-$k$ recommendation list generated by a model for a sample in $T$, $r^i$ refers to the $i$-th recommended item in the list, and $H[r^i]$ is an indicator function that returns 1 if $r^i$ is labeled as healthy, and 0 otherwise. Notably, HP@$k$ treats all positions among the list equally. In contrast, HO@$k$ emphasizes the ranking importance by assigning higher weights to healthy content appearing earlier in the list, using a discounting scheme of $[1, \frac{1}{2}, \frac{1}{3}, \dots, \frac{1}{k}]$. The result is normalized by $\sum_{i=1}^{k} \frac{1}{i}$ to ensure that the final score remains within the range [0, 1]. For both HP@$k$ and HO@$k$, a higher value indicates a stronger ability to prioritize healthy content. Further discussion of the proposed metrics can be found in Appendix~\ref{app_metric}.

\section{Empirical Findings}\label{sec_warning}
Based on the constructed benchmarks, we further examine current recommendation methods from the harmfulness perspective. Specifically, we focus on answering three key questions: (a) \textit{Does harmful content exist in the recommendation data?} (b) \textit{How does harmful content influence user behavior?} and (c) \textit{Do current recommendation techniques facilitate the propagation of harmful content?}

\subsection{Warning 1: Harmful content widely exists in recommendation data.}

\fig~\ref{findings} (a) reports the proportion of harmful versus healthy content in three constructed dataset. Alarmingly, we observe that \textbf{more than 50\% of the content in each dataset is labeled as harmful}. This indicates that harmful content is not a marginal phenomenon but instead dominates real-world data, posing serious risks by undermining individual values and threatening broader societal stability. Although major content service providers have established clear regulations to prohibit the production and dissemination of harmful material, the massive volume of content generated daily renders comprehensive monitoring unfeasible. As a result, harmful content inevitably continues to persist and circulate within these platforms. This finding underscores the pressing importance of considering content harmfulness in the modern digital era.

\subsection{Warning 2: Users tend to engage with harmful content.}

\begin{table*}[t]
\tabcolsep 0.1in 
  \centering
    \caption{Harmfulness performance (\%) of various popular recommendation methods.}
    \begin{tabular}{c |ccc |ccc| ccc}  
    \toprule
    
    \multirow{2}{*}{Method}  & \multicolumn{3}{c}{TN} & \multicolumn{3}{c}{QB}    &\multicolumn{3}{c}{EB-NeRD} \cr
    \cmidrule(lr){2-4}\cmidrule(lr){5-7} \cmidrule(lr){8-10}
    &HP@1    & HP@5 & HO@5  &HP@1    & HP@5 & HO@5    &HP@1    & HP@5   & HO@5 \cr
    \midrule 
    GRU4Rec~\cite{GRU4Rec}  &47.77  &46.68    &45.77  &50.62  &51.18   &48.69  &48.31  &49.89   &49.46 \\
    NARM~\cite{NARM}    &47.60  &47.42    &48.41  &50.89  &50.03   &51.00   &49.32  &49.52  &49.17 \\
    SASRec~\cite{SASRec}  &45.58  &46.71    &46.35  &47.99  &47.12  &47.31    &\textbf{50.55}  &49.84 &\textbf{50.36} \\
    BERT4Rec~\cite{BERT4Rec}    &47.13  &46.65    &48.38  &\textbf{51.01}  &\textbf{51.77} &51.05  &48.17  &50.16   &49.36 \\
    SR-GNN~\cite{SR-GNN}  &\textbf{50.05}  &\textbf{49.19}  &\textbf{50.07}  &46.75  &49.39   &48.69  &43.56  &44.19  &44.29 \\
    LESSR~\cite{LESSR}   &46.62  &46.34 &46.57  &47.54  &49.66   &49.58  &50.10  &\textbf{50.52}  &50.34 \\
    PLM-NR~\cite{Wu@SIGIR2021}  &42.24  &46.45   &45.24  &34.91  &43.70   &41.07     &48.32  &48.43    &48.36 \\
    MSGIFSR~\cite{guo@WSDM2022} &47.54  &48.38    &48.28     &50.21  &51.18    &\textbf{51.06}     &50.31  &50.46    &50.24 \\
    Atten-Mixer~\cite{Zhang@WSDM2023} &41.62  &43.41    &42.93     &45.61  &49.06    &47.28     &48.99  &50.22    &49.52 \\
    Prompt4NR~\cite{Zhang@SIGIR2023}   &46.49  &47.64   &47.30     &49.52  &49.18    &49.22     &43.27  &44.19    &44.06 \\
    BSARec~\cite{Shin@AAAI2024}   &44.69  &48.72    &47.76     &38.08  &44.43    &42.06     &46.25  &48.18    &47.39 \\
    DIMO~\cite{DIMO}    &49.07  &48.66   &48.44     &48.32  &47.21    &47.30     &47.35  &47.01    &47.05 \\
    Oracle4Rec~\cite{Xia@WSDM2025}  &43.25  &43.35    &43.71  &42.07  &43.18   &43.69  &46.16  &46.24   &47.96 \\

    \bottomrule
    \end{tabular}
    \label{all_health}
\end{table*}

We further analyze user behaviors with respect to harmful and healthy content from two complementary perspectives. On the one hand, we report the average number of user interactions per item for harmful and healthy content in \fig~\ref{findings} (b), calculated as the total interactions on harmful (or healthy) items divided by the number of harmful (or healthy) items. This metric reflects the relative engagement intensity associated with each content type. On the other hand, we identify users who have interacted with at least one harmful item in their historical behavior sequences and calculate the proportion of such users among the entire user population. This analysis provides insight into the prevalence of exposure to harmful content at the user level.

As illustrated in \fig~\ref{findings} (b), \textbf{harmful items consistently attract more user interactions than healthy items}. Specifically, the average number of interactions per item is 37.84 vs. 35.32 in TN, 24.60 vs. 20.37 in QB, and 338.81 vs. 306.33 in EB-NeRD for harmful and healthy content, respectively. Moreover, \textbf{all users (\ie 100\%) in TN, QB, and EB-NeRD have interacted with at least one harmful item in their historical activities}. These observations suggest that users exhibit a stronger tendency to engage with harmful content than with healthy content. This phenomenon is not entirely unexpected, as harmful content, such as superstition, gender discrimination, offensive language, pornography, and misinformation, is often more provocative and attention-grabbing, making it more likely to elicit user interactions. Importantly, this tendency suggests that user behavior may inadvertently contribute to the amplification of harmful content, underscoring the necessity of examining information propagation from a harmfulness-aware perspective.
In addition, the average number of interactions per item varies substantially across three datasets, \ie around 30, 20, and 300 for TN, QB, and EB-NeRD, respectively. As shown in Table~\ref{statistics}, EB-NeRD contains over two million interactions, substantially more than TN and QB (around 100K each), resulting in much denser item interactions.

\subsection{Warning 3: Current recommendation methods indeed facilitate the propagation of harmful content.}

In this section, we select 13 representative recommendation models to systematically investigate the harmfulness of RSs' outputs. 
These models span a broad range of architectures and modeling paradigms, including Recurrent Neural Networks (GRU4Rec~\cite{GRU4Rec}, NARM~\cite{NARM}), Attention mechanism (SASRec~\cite{SASRec}, Atten-Mixer~\cite{Zhang@WSDM2023}, BSARec~\cite{Shin@AAAI2024}, DIMO~\cite{DIMO}, Oracle4Rec~\cite{Xia@WSDM2025}), Transformer structures (BERT4Rec~\cite{BERT4Rec}, BSARec~\cite{Shin@AAAI2024}), Graph Neural Networks (SR-GNN~\cite{SR-GNN}, LESSR~\cite{LESSR}, MSGIFSR~\cite{guo@WSDM2022}), as well as large language models (Prompt4NR~\cite{Zhang@SIGIR2023}). Collectively, these methods represent major advances in sequential recommendation from 2016 to now. For each model, we process item text using a text encoder (\eg BERT~\cite{BERT}) and utilize the resulting text embeddings to initialize the item representations across all models. 
Each method is executed five times with different random seeds, and the averaged results are reported to ensure the reliability of the evaluation. 
Detailed descriptions of these methods are provided in Appendix~\ref{all_models}.

We report their performance under the proposed HP@$k$ and HO@$k$ metrics ($k$ $\in$ \{1, 5\}) in Table~\ref{all_health}. As shown, the proportion of healthy content in recommendation results is below 50\% across almost all models, indicating that \textbf{more than half of the recommended items are harmful}, involving content such as violence, hate speech, and pornography. As recommender systems largely determine the content users are exposed to and engage with, this finding reveals a concerning reality: current recommendation methods indeed facilitate the propagation of harmful content. This outcome is also intuitive, as current methods are primarily designed to capture user interest and maximize engagement, without accounting for content harmfulness. Consequently, the harmfulness of recommendation outputs tends to reflect the underlying data distribution, which is dominated by harmful content, as discussed in Warning 1. Moreover, this observation accounts for the Warning 2, suggesting that extensive exposure to harmful content may encourage more frequent user engagement with such content. This observation (Warning 3) underscores the urgent need for the community to systematically investigate recommender systems from a harmfulness perspective, a critical dimension that has been largely overlooked in existing research.

\section{Problem Formulation}\label{problemform}
Sequential recommendation aims to predict the next item a user is likely to interact with based on her historical behaviors~\cite{NARM,Cui@NIPS2025,Li@WWW2026}. Let $\mathcal{I}$ and $\mathcal{U}$ denote the sets of items and users, respectively, where $n = |\mathcal{I}|$ represents the total number of items. For a user $u\in\mathcal{U}$, $\mathcal{S}=[x_1,x_2,\cdots,x_t]$ denotes her historical interaction sequence, where $x_i\in\mathcal{I}$ and $t$ is the sequence length. In general, existing recommendation models learn item and user embeddings, whose similarities determine the recommendation results. Specifically, an item $x_i$ is represented by an embedding $\mathbf{v}_i\in\mathbb{R}^{d}$ ($d$ is the embedding dimension) obtained by encoding its textual features with a language model $textEncoder(\cdot)$, \eg BERT~\cite{BERT}. Then, a sequence encoder $seqEncoder(\cdot)$, \eg an attention-based encoder~\cite{SASRec,BERT4Rec}, operates on $\mathcal{S}$ to obtain the user embedding $\mathbf{u}\in\mathbb{R}^{d}$. The relevance score between user $u$ and a candidate item $x_i$ is estimated as $y'_i=\langle\mathbf{u},\mathbf{v}_i\rangle$, where $\langle\cdot,\cdot\rangle$ denotes a similarity function, such as the dot product. The top-$k$ items with the highest scores are then selected to form the recommendation list $r=[x'_1,x'_2,\cdots,x'_k]$. Typically, the sequential recommendation task is optimized using the cross-entropy loss as follows:
\begin{align}
    \mathcal{L}_{rec} = - \sum^n_{i=1} p_i \log (y^{\prime}_i) + (1-p_i)\log(1-y^{\prime}_i),
\end{align}
where $p_i$ is the ground truth signifying whether a user interacts with item $x_i$. Note that, for each recommended item $x'_i$, existing methods primarily focus on whether it aligns with the user's interests, while overlooking its potential harmfulness. As revealed in Section~\ref{sec_warning}, this limitation leads to the widespread presence of harmful content in recommendation lists. Therefore, this work aims to reduce the exposure of harmful content in recommendation lists, thereby mitigating its propagation.

\section{Mitigation Framework}

\subsection{Harmfulness-aware item representation learning.} As discussed in Section~\ref{sec:health_label}, the harmfulness reason provides a clear and informative signal indicating content harmfulness. Thus, we leverage the language model $textEncoder(*)$ to convert the reason into a dense vector $\mathbf{v}_i^{h} \in \mathbb{R}^{d}$, serving as the harmfulness-aware embedding for item $x_i$. 
It is worth noting that due to the inherent complexity of natural language, especially the use of implicit expressions such as metaphors and rhetoric, the original content-based representation $\mathbf{v}_i$ fails to fully capture an item's harmfulness. By incorporating the generated harmfulness reasons, our \baby framework can better recognize and represent the harmfulness of content, enabling more accurate distinction between healthy and harmful items.

\subsection{Harmfulness-aware user representation learning.} 

Self-attention has been widely adopted in recommendation tasks due to its strong ability to capture item transition patterns within user behavior sequences~\cite{Zhu@AAAI2019, BERT4Rec}. However, standard implementations of self-attention are unaware of the harmfulness of item content, resulting in harmfulness-agnostic user representations. To overcome this limitation, we present a harmfulness-aware self-attention mechanism to obtain harmfulness-aware user embeddings.

\begin{figure}
  \centering
  \includegraphics[width=0.9\linewidth]{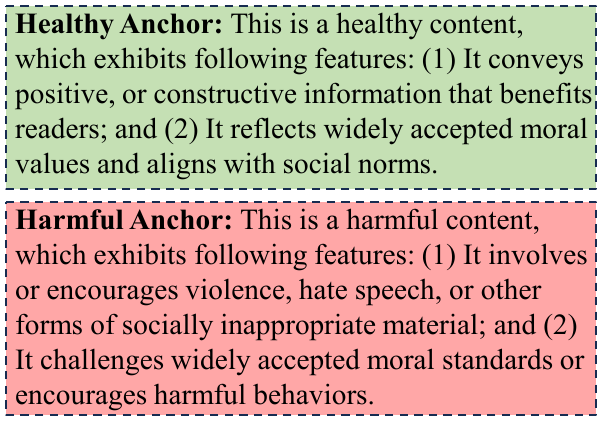}
  \caption{Healthy and harmful anchors.}\label{anchor}
\end{figure}

Specifically, we construct a synthetic healthy content, named healthy anchor $x^{+}$, as shown in Figure~\ref{anchor}, which serves as a clean and fully healthy signal. The healthy anchor is generated by reversing the harmful semantics specified in our harmful content definition.
We then employ the pre-trained language model $textEncoder(*)$ to convert the healthy anchor $x^{+}$ into a dense vector  $\mathbf{v}^{+} \in \mathbb{R}^d$. By replicating this vector, we form a healthy anchor sequence $\mathbf{E}^+$ = [$\mathbf{v}^{+}$, $\mathbf{v}^{+}$, $\cdots$, $\mathbf{v}^{+}$]. Formally, for a sequence of harmfulness-aware item embeddings $\mathbf{E}_h$ = [$\mathbf{v}^h_1$, $\mathbf{v}^h_2$, $\cdots$, $\mathbf{v}^h_t$], the harmfulness-aware self-attention mechanism operates as follows:
\begin{align}
    H &= softmax(\frac{Q^TK}{\sqrt{d}})V, \\
    Q &= W^Q\mathbf{E}^{+}, K= W^K\mathbf{E}_h, V= W^V\mathbf{E}_h,
\end{align}
where $W^Q$, $W^K$ and $W^V \in \mathbb{R}^{d \times d}$ are learnable matrices used to map the inputs into query, key and value spaces, respectively, and $H$ = [$\mathbf{h}_1, \mathbf{h}_2, \cdots, \mathbf{h}_t$] is the output. Afterwards, the harmfulness-aware user embedding is derived by an average operation as $\mathbf{u}^h$= $\frac{1}{t}\sum_{k=1}^{t} \mathbf{h}_k$. In this mechanism, the synthetic healthy anchors $\mathbf{E}^+$ are used as the Query to compute attention weights for each history item. By doing so, healthy content receives greater attention weights compared to harmful content, thereby emphasizing healthy signals while suppressing the harmful ones in user representation learning. 
Thus, this attention allocation encourages user embeddings to lean toward healthy content.

In addition, we explicitly drive user embeddings to be closer to healthy content while pushing them away from harmful content. Inspired by contrastive learning~\cite{chen@ICML2020}, we formulate this objective by minimizing the following contrastive loss,
\begin{align}
    \mathcal{L}_{health} &= -\frac{<\mathbf{u}^h, \mathbf{v}^+>}{<\mathbf{u}^h, \mathbf{v}^->},
\end{align}
where $\mathbf{v}^- \in \mathbb{R}^d$ denotes the embedding of the synthetic harmful anchor, \ie a clean and explicitly harmful signal, as illustrated in \fig~\ref{anchor}.

\subsection{Harmful Content Mitigation.} As discussed in Section~\ref{problemform}, current recommendation methods typically capture user interest via $y^{\prime}_i$ = $<\mathbf{u}$, $\mathbf{v}_i>$ for recommendations, while failing to account for content harmfulness. To address this limitation, we incorporate content harmfulness as an additional signal into the recommendation scoring function as follows,
\begin{align}
    y_i &= <\mathbf{u}, \mathbf{v}_i> + \lambda(<\mathbf{u}^h, \mathbf{v}^h_i> + <\mathbf{v}^h_i, \mathbf{v}^+>),\label{h_pro}
\end{align}
where $\lambda$ is a hyper-parameter that balances user interest and content harmfulness. In Eq.(\ref{h_pro}), the first term captures the user's interest in the item, while the second term accounts for content harmfulness. Specifically, $\mathbf{v}^h_i$ explicitly captures the harmfulness of an item. Both $<\mathbf{u}^h, \mathbf{v}^h_i>$ and $<\mathbf{v}^h_i, \mathbf{v}^+>$ work together to enlarge the scores of healthy content while suppressing those of harmful ones, as $\mathbf{u}^h$ is guided toward healthy content and $\mathbf{v}^+$ represents a clean, fully healthy signal. This enhanced scoring mechanism enables \baby to encourage healthy content exposure. Finally, the proposed \baby framework is trained as follows:
\begin{align}
    \mathcal{L} &= \mathcal{L}_{rec} + \mathcal{L}_{health},
\end{align}
where $\mathcal{L}_{rec}$ is responsible for learning interest-driven representations including $\mathbf{u}$ and $\mathbf{v}_i$ in base models. In contrast, $\mathcal{L}_{health}$ focuses on harmfulness-related representation learning, such as $\mathbf{u}^h$ and $\mathbf{v}_i^h$. 

\subsection{Discussion.}
To incorporate content harmfulness into recommendation, \baby customizes several techniques, including injecting explicit harmfulness signals into item embeddings, introducing a harmfulness-aware self-attention mechanism, and adopting modified contrastive learning. These components work together to enable harmfulness-aware recommendation. Notably, our \baby framework operates without requiring any modifications to base models. This flexibility allows \baby to be seamlessly integrated into a wide range of existing neural recommenders, enhancing their ability to mitigate harmful content propagation.

\section{Experiments}
\subsection{Experiment Setups}\label{sec_setup}

\subsubsection{Research Questions}
Extensive experiments are conducted on the constructed benchmarks to evaluate the performance of the proposed \baby and all baselines, with the goal of addressing the following research questions: \textbf{RQ1}: How does our \baby perform compared with existing methods, particularly in terms of mitigating recommendation harmfulness? (ref.Section~\ref{sec_overall}) \textbf{RQ2}: Do the proposed components contribute positively to \baby's performance? (ref. Section~\ref{sec_abl}) \textbf{RQ3}: How do key hyperparameters influence the performance of \baby? (ref. Section~\ref{sec_hyper}) \textbf{RQ4}: What is the efficiency of the \baby? (ref. Section~\ref{sec_efficiency}) \textbf{RQ5}: How well dose the proposed \baby work in the real-world instance?  (ref. Section~\ref{sec_case})

\begin{table*}[t]
\tabcolsep 0.1in 
  \centering
    \caption{Overall performance (\%) of \baby framework on several base recommendation models. Statistical significance (*) is determined by the t-test ($p < 0.01$).}
    \begin{tabular}{c |ccc |ccc| ccc}  
    \toprule
    
    TN  &GRU4Rec    &\baby  &$imprv.$ &NARM    &\baby  &$imprv.$ &BERT4Rec    &\baby  &$imprv.$ \cr
    \midrule 
    Prec@1 &\textbf{14.36}  &14.05  &-2.16\%  &\textbf{15.76}  &15.39  &-2.35\%  &10.78  &\textbf{10.94}  &+1.48\% \\
    Prec@5 &\textbf{27.39}  &26.96  &-1.55\%  &\textbf{31.37}  &30.89  &-1.53\%  &\textbf{27.64}  &27.46  &-0.65\% \\
    MRR@5  &\textbf{19.05}  &18.84  &-1.10\%  &\textbf{21.12}  &20.73  &-1.87\%  &16.62  &\textbf{16.71}  &+0.54\% \\
    \rowcolor{healthcolor} HP@1 &47.77  &\textbf{63.45}  &+32.84\%  &47.60  &\textbf{66.72}  &+40.17\%  &47.13  &\textbf{66.29}  &+40.65\% \\
    \rowcolor{healthcolor} HP@5 &46.68  &\textbf{62.82}  &+34.59\%  &47.42  &\textbf{64.10}  &+35.18\%  &46.65  &\textbf{62.02}  &+32.95\% \\
    \rowcolor{healthcolor} HO@5 &45.77  &\textbf{62.54}  &+36.65\%  &48.41  &\textbf{65.22}  &+34.72\%  &48.38  &\textbf{64.19}  &+32.67\% \\
    \midrule

    QB  &GRU4Rec    &\baby  &$imprv.$ &NARM    &\baby  &$imprv.$ &BERT4Rec    &\baby  &$imprv.$ \cr
    \midrule 
    Prec@1 &70.10  &\textbf{70.86}  &+1.08\%  &\textbf{72.35}  &71.89  &-0.64\%  &\textbf{69.66}  &69.50  &-0.23\% \\
    Prec@5 &75.30  &\textbf{76.52}  &+1.62\%  &\textbf{78.23}  &77.96  &-0.35\%  &\textbf{88.02}  &87.57  &-0.51\% \\
    MRR@5 &72.01  &\textbf{73.30}  &+1.79\%  &\textbf{74.67}  &74.55  &-0.16\%  &\textbf{77.01}  &76.62  &-0.51\% \\
    \rowcolor{healthcolor} HP@1 &50.62  &\textbf{62.55}  &+23.57\%  &50.89  &\textbf{64.91}  &+27.55\%  &51.01  &\textbf{64.13}  &+25.72\% \\
    \rowcolor{healthcolor} HP@5 &51.18  &\textbf{64.58}  &+26.18\%  &50.03  &\textbf{77.05}  &+54.01\%  &51.77  &\textbf{68.17}  &+31.68\% \\
    \rowcolor{healthcolor} HO@5 &48.69  &\textbf{73.19}  &+50.32\%  &51.00  &\textbf{73.64}  &+44.39\%  &51.05  &\textbf{67.06}  &+31.36\% \\
    \midrule

    EB-NeRD  &GRU4Rec    &\baby  &$imprv.$ &NARM    &\baby  &$imprv.$ &BERT4Rec    &\baby  &$imprv.$ \cr
    \midrule 
    Prec@1 &2.79  &\textbf{2.91}  &+4.31\%  &0.85  &\textbf{0.93}  &+9.41\%  &\textbf{11.58}  &11.29  &-2.50\% \\
    Prec@5 &\textbf{8.33}  &8.24  &-1.02\%  &2.99  &\textbf{3.40}  &+13.71\%  &\textbf{36.72}  &36.42  &-0.82\% \\
    MRR@5 &4.82  &\textbf{4.86}  &+0.93\%  &1.64  &\textbf{1.74}  &+6.10\%  &19.66  &\textbf{19.94}  &+1.42\% \\
    \rowcolor{healthcolor} HP@1 &48.31  &\textbf{64.76}  &+34.05\%  &49.32  &\textbf{58.32}  &+18.25\%  &48.17  &\textbf{58.21}  &+20.84\% \\
    \rowcolor{healthcolor} HP@5 &49.89  &\textbf{63.23}  &+26.74\%  &49.52  &\textbf{55.85}  &+12.78\%  &50.16  &\textbf{57.40}  &+14.42\% \\
    \rowcolor{healthcolor} HO@5 &49.46  &\textbf{64.69}  &+30.79\%  &49.17  &\textbf{57.15}  &+16.23\%  &49.36  &\textbf{56.70}  &+14.87\% \\

    \bottomrule
    \end{tabular}
    \label{performance}
\end{table*}

\subsubsection{Datasets} Enriched \textbf{TN}, \textbf{QB} and \textbf{EB-NeRD} are employed to examine the performance of our \baby and baselines from both accuracy and harmfulness aspects.

\subsubsection{Evaluation Metrics} 
We evaluate the performance of RS from both accuracy and harmfulness perspectives. For accuracy evaluation, we adopt two widely used metrics, \textbf{Prec@k} (Precision) and \textbf{MRR@k} (Mean Reciprocal Rank). As to harmfulness assessment, we employ the newly proposed \textbf{HP@k} and \textbf{HO@k}. In this work, k is set to 1 and 5. Note that for all these metrics, higher values indicate better performance, \ie more accurate or healthier recommendations.

\subsubsection{Base models} We instantiate our \baby framework on three representative models: GRU4Rec~\cite{GRU4Rec}, NARM~\cite{NARM}, and BERT4Rec~\cite{BERT4Rec}. These models are widely regarded as foundational architectures and standard benchmarks in the recommendation community. Therefore, demonstrating the effectiveness of \baby on these backbones suggests its strong potential for generalization to a broad range of recommendation approaches.

\subsubsection{Implementation details} The pre-trained text encoder $textEncoder(*)$ is instantiated with BERT~\cite{BERT}, and dot-product is used as the similarity function $<*>$. 
To ensure a fair comparison, we fix the embedding dimension at $d=100$, use the Adam optimizer with a learning rate of $1 \times 10^{-3}$, and set the mini-batch size to 512 for all base models. All hyper-parameters are tuned via grid search based on models' performance in terms of Prec@5 on the validation set. The base models and their enhanced versions with \baby share identical experimental settings. The balancing coefficient $\lambda$ in \baby is searched from \{0.001, 0.01, 0.1, 1\}. We run each method 5 times independently, and use the average performance as the final results. Further details of experiment setups can be found in Appendix~\ref{app_setup}.
All resources of this work are available online at \url{https://github.com/Zhang-xiaokun/MitiHarm}.

\subsection{Recommendation Harmfulness and Accuracy of \baby}\label{sec_overall}

As shown in Table~\ref{performance}, for each group, we present results for the base model, its enhanced version with \baby, as well as improvement achieved by \baby. The best results in each case are highlighted in bold. From the table, we can draw the following observations:

\subsubsection{Harmfulness comparison} Our proposed \baby framework demonstrates a substantial enhancement in terms of harmfulness across various models. For instance, when applied to NARM, \baby achieves a 54.01\% improvement on HP@5 metric in QB, indicating its effectiveness in improving the proportion of healthy content in recommendations. Similarly, in terms of HO@5 in QB, \baby delivers a notable 50.32\% gain with GRU4Rec, showcasing its ability to push healthy content toward the top of the recommendation list. These improvements suggest that \baby increases the presence of healthy content and strategically prioritizes its visibility to users. By doing so, \baby can effectively mitigate the propagation of harmful content, achieving more responsible recommendations and contributing to a positive societal impact.

\subsubsection{Accuracy comparison} While incorporating \baby leads to slight accuracy drops in some cases, the decline is marginal, typically less than 0.5\% in absolute terms. Given the substantial gains it brings in harmfulness mitigation, such trade-offs are acceptable. More interestingly, \baby can improve accuracy in certain cases, \eg GRU4Rec on QB. This suggests that \baby does not merely trade accuracy for harmfulness. It can enhance both dimensions simultaneously. Such a dual benefit contributes to the development of both socially responsible and user-satisfied recommendations.

\begin{figure*}[t]
  \centering
  \includegraphics[width=0.85\linewidth]{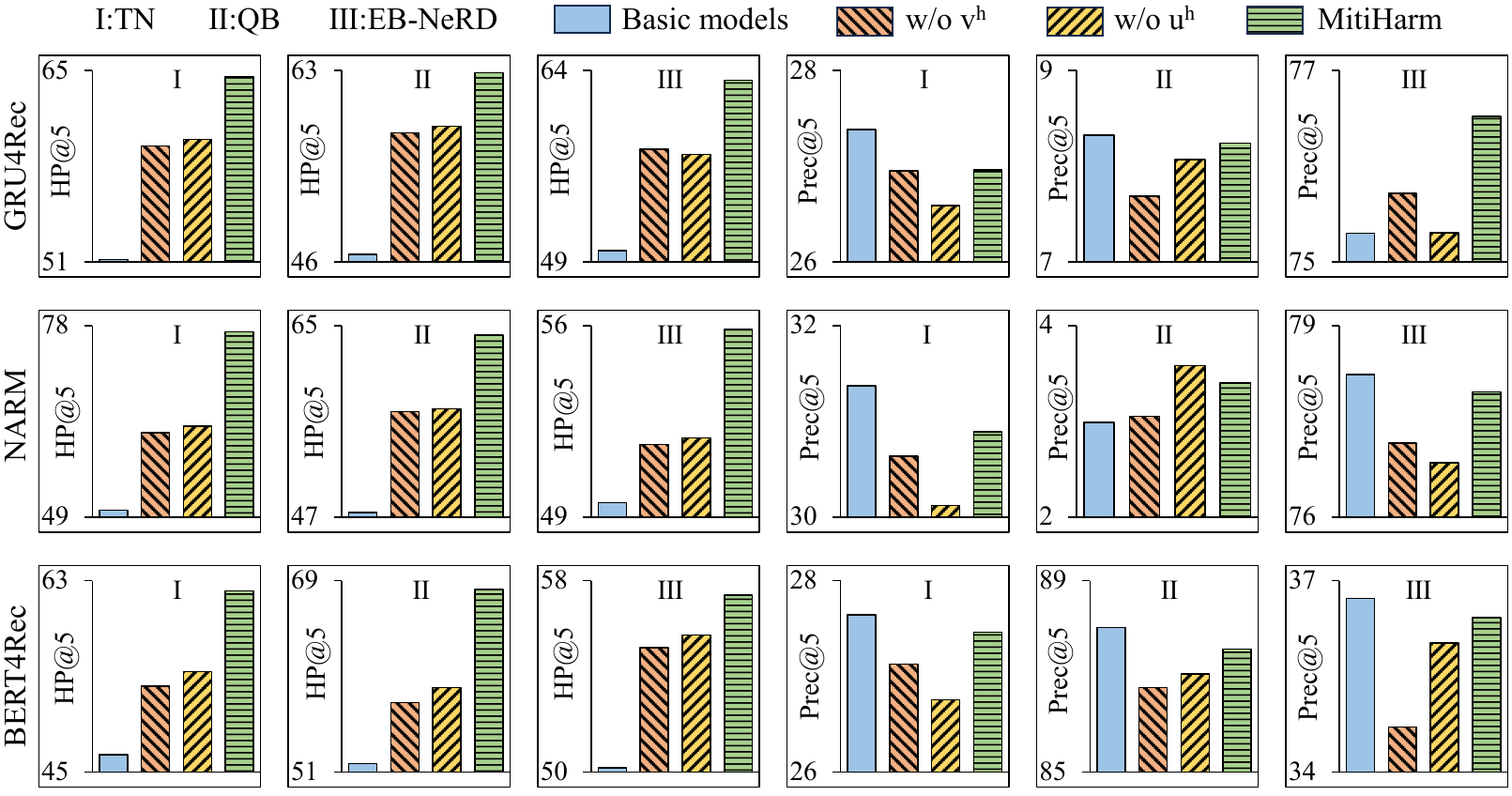}
  \caption{Ablation study on harmfulness-aware item and user representation learning.}\label{ablation}
\end{figure*}

\subsubsection{Applicability of \baby} Our \baby consistently mitigates recommendation harmfulness for all base models across all datasets, while maintaining impressive prediction accuracy. Its ability to deliver stable and effective performance confirms its generic and model-agnostic nature, making it a practical and scalable solution for promoting the propagation of healthy content under the recommendation scenario.

\subsection{Ablation Study}\label{sec_abl}

In this section, we analyze the effectiveness of harmfulness-aware item and user representation learning in \baby. Following two variants of \baby are designed to conduct the ablation study: (\romannumeral1) w/o $\mathbf{v}^h$ replaces the harmfulness-aware item embeddings $\mathbf{v}^h$ with the original item embeddings $\mathbf{v}$ in $\mathbf{E}_h$ and Eq.(\ref{h_pro}), discarding the explicit capture of item harmfulness; and (\romannumeral2) w/o $\mathbf{u}^h$ removes the harmfulness-aware self-attention and contrastive learning from \baby. Instead, it only adopts user embeddings from the base models in Eq.(\ref{h_pro}). Following observations can be made from \fig~\ref{ablation}:

First, both variants are defeated by the complete \baby in HP@5 metric, demonstrating the effectiveness of individual components of \baby in terms of recommendation harmfulness. By injecting harmfulness-related information into either item or user representations, these components are able to capture content harmfulness. Second, the variant w/o $\mathbf{v}^h$ achieves healthier performance over base model, demonstrating the harmfulness-aware capability of our proposed self-attention mechanism, where the healthy anchor serves as the query to guide user embeddings toward healthy content. Finally, both variants lead to an accuracy decrease compared to the complete \baby. Different from \baby that clearly disentangles the modeling for user interest and content harmfulness, these variants mix these two factors, making it more challenging to generate accurate predictions. Specifically, the variant w/o $\mathbf{u}^h$ combines the harmfulness-aware item embeddings with the original user embeddings to generate recommendations. As to the variant w/o $\mathbf{v}^h$, it indiscriminately merges harmfulness-aware user embeddings with original item embeddings to do so.

\subsection{Hyper-parameter study}\label{sec_hyper}
\begin{figure*}[t]
  \centering
  \includegraphics[width=0.99\linewidth]{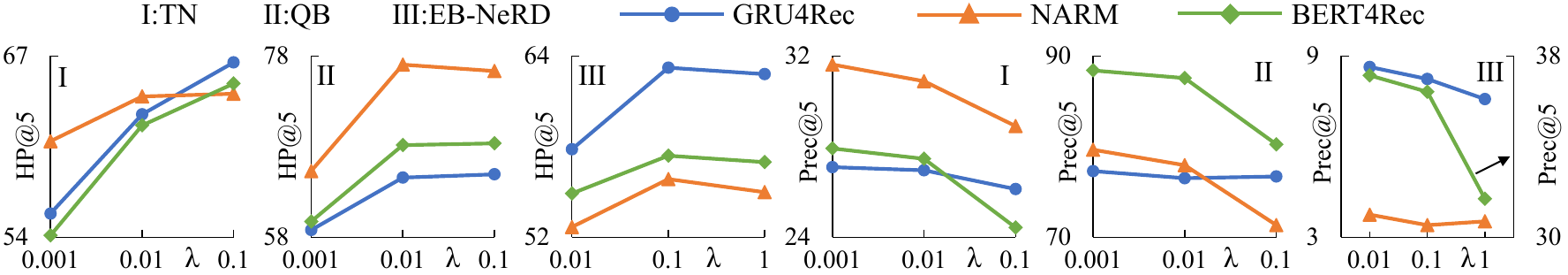}
  \caption{The impact of $\lambda$ on the performance of \baby in terms of harmfulness and accuracy.}\label{hyper}
\end{figure*}

\begin{table*}[t]
\tabcolsep 0.1in 
  \centering
    \caption{Time efficiency of the proposed framework \baby.}
    \begin{tabular}{c  |cc |cc |cc}  
    \toprule  
    TN  &GRU4Rec &\baby   &NARM &\baby  &BERT4Rec &\baby   \\
    \midrule
    Converge Epochs &68  &70  &74  &72  &69  &72 \\
    Time Per Epoch &0.367s  &0.389s  &0.324s  &0.367s  &2.559s  &2.693s \\
    Inference Time  &0.154s  &0.162s  &0.158s  &0.164s  &0.288s  &0.303s \\
    \midrule

   QB  &GRU4Rec &\baby   &NARM &\baby  &BERT4Rec &\baby   \\
   \midrule
    Converge Epochs &68  &67  &79  &77  &71  &80\\
    Time Per Epoch &0.224s  &0.332s  &0.274s  &0.286s  &2.838s  &3.302s\\
    Inference Time  &0.157s  &0.167s  &0.160s  &0.238s  &0.384s  &0.400s\\
    \midrule

    EB-NeRD  &GRU4Rec &\baby   &NARM &\baby  &BERT4Rec &\baby   \\
    \midrule
    Converge Epochs &71  &68  &60  &65  &61  &65 \\
    Time Per Epoch &0.598s  &2.075s  &2.692s  &2.884s  &60.686s  &71.679s\\
    Inference Time  &0.217s  &0.394s  &0.221s  &0.236s  &8.288s  &8.571s\\
    
	\bottomrule
    \end{tabular}
    \label{efficiency}
\end{table*}

In our \baby framework, the key hyperparameter is $\lambda$ in Eq.(\ref{h_pro}), which balances user interest and content harmfulness when generating recommendations. As shown in \fig~\ref{hyper}, we plot the performance trends of \baby with respect to $\lambda$ in terms of both harmfulness and accuracy.
As observed in the figure, the models' capability to recommend healthy content improves as $\lambda$ increases. This is intuitive, as a larger $\lambda$ drives the model to prioritize recommending more healthy content. 
Regarding accuracy, it initially decreases slightly, then experiences a sharp decline as $\lambda$ continues to rise. This indicates that \baby is effective at balancing accuracy and harmfulness up to a certain point, however, once $\lambda$ exceeds a threshold, this balance begins to deteriorate. This phenomenon can be explained by the findings in Section~\ref{sec_warning}: (a) harmful content constitutes a large proportion of real-world data; and (b) users tend to interact more frequently with harmful content than with healthy content. Thus, overly prioritizing healthy content may deviate from user behavior patterns, leading to a noticeable decline in recommendation accuracy. 
Based on this, we empirically select the optimal $\lambda$, ensuring that our \baby mitigates the harmfulness of recommendations while still effectively capturing user interest.

\subsection{Efficiency Analysis}\label{sec_efficiency}

To evaluate the efficiency of the proposed \baby framework, we record the performance of both base models and their \baby-enhanced versions under three key efficiency metrics, including the number of convergence epochs (Converge Epochs), Time Per Epoch, and Inference Time. As shown in Table~\ref{efficiency}, the base models and their \baby versions converge in a similar number of epochs. In some cases, such as GRU4Rec on QB and EB-NeRD, and NARM on TN and QB, \baby even converges slightly faster than its base model. This observation indicates that highlighting content harmfulness during conducting recommendations does not introduce additional optimization difficulty or training overhead. 
Regarding training time, \baby introduces a modest increase in training time over its base models. For example, on the TN dataset, the additional time is approximately 0.02 seconds for GRU4Rec, 0.04 seconds for NARM, and 0.13 seconds for BERT4Rec per training step. This time increase can be attributed to the harmfulness-aware self-attention mechanism and the additional contrastive loss introduced by \baby. Similarly, there is a small increase in inference time, \eg typically less than 0.05 seconds in most cases. It is also intuitive, as \baby considers content harmfulness that base models ignore during recommendation generation. 
Nevertheless, the significant improvements in harmfulness metrics make this trade-off acceptable. In summary, the overall efficiency of \baby remains high, making it a lightweight and scalable framework suitable for practical applications.

\subsection{Case Study}\label{sec_case}

\begin{figure*}[t]
  \centering
  \includegraphics[width=0.95\linewidth]{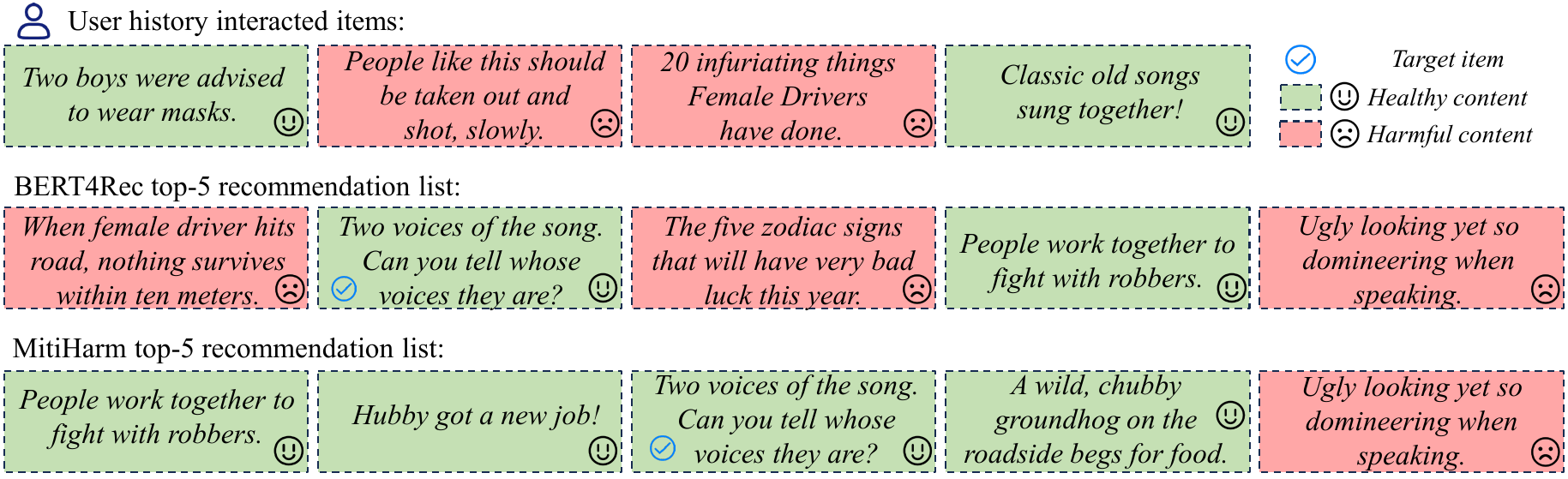}
  \caption{Case study of \baby's recommendations.}\label{case}
\end{figure*}

In order to intuitively scrutinize \baby's performance in handling the recommendation task, we present a case from TN dataset in~\fig~\ref{case}, which includes a user’s interaction sequence alongside the recommendation lists generated by BERT4Rec and its enhanced version with \baby. From this example, we observe the presence of various harmful content in real-world platforms, such as violence, gender discrimination and superstition. Even worse, current methods like BERT4Rec tend to spread such harmful content, as most of its recommended items fall into this category. In contrast, our \baby mitigates the propagation of harmful content, with the majority of its recommendations being healthy. Besides, \baby hits the target item of user interest, demonstrating its ability to align with user preferences.

\section{Related Work}
\subsection{Accuracy-oriented recommendation}

Current methods in recommender systems are dedicated to improving prediction accuracy via capturing users' interest, involving the utilization of various techniques, such as recurrent neural networks~\cite{GRU4Rec}, attention mechanism~\cite{SASRec, BERT4Rec}, graph neural networks~\cite{SR-GNN, CoHHN}, reinforcement learning~\cite{Alvaro@ICML2024}, and self-supervised learning~\cite{Jiang@SIGIR2023, Jiang@CIKM2023}. Additionally, recent efforts attempt to incorporate LLMs into the task, using LLMs to enhance the understanding for item content~\cite{Zhang@SIGIR2023, Gao@WWW2024, Kong@NIPS2024, Kim@ICLR2026}. However, all of these methods focus on catering to user interest, usually overlooking the harmfulness of the recommended content. 

\subsection{Quality-enhanced recommendation}
In addition to accuracy-oriented works, some efforts have attempted to enhance the quality of content recommendation. Here, the ``quality'' covers a wide range of aspects within recommendations, such as eliminating fake news~\cite{Wang@KDD2024}, improving recommendation fairness~\cite{Chen@NIPS2024,Greenwood@NIPS2024} or interpretability~\cite{Lyu@ICDE2021}, addressing the biased popularity of items~\cite{Gupta@ICML2021, Ashutosh@SIGIR2023, Liao@TKDE2026, Abdulla@WSDM2026}, mitigating unwanted recommendations~\cite{Giovanni@RecSys2025}, avoiding radicalization~\cite{Ribeiro@FAT2020}, as well as increasing business revenue~\cite{Huang@WWW2024, Chen@KDD2024} or recommendation diversity~\cite{Li@KDD2024, Itay@ICML2023}. Moreover, some methods aim to recommend high-quality items to users. Typically, they rely on explicit or implicit user feedback to assess an item's quality, such as scroll information~\cite{Lagun@SIGIR2014}, ratings~\cite{Xin@WSDM2011, Zhang@RecSys2018}, and dwell time~\cite{Lu@WWW2019, Wang@RecSys2023}. More recently, \cite{Chee@KDD2024} highlights the existence of harmful content in the movie domain. However, their analysis assumes that recommender systems never recommend harmful content, an assumption that is inconsistent with real-world observations and the empirical findings of our study.

\section{Conclusion}
In this work, we take an initial step toward investigating recommender systems from the harmfulness perspective, an important yet under-explored dimension. We first construct harmfulness-annotated datasets through a LLM-human collaborative annotation paradigm and introduce two harmfulness-related metrics, HP@$k$ and HO@$k$, to quantitatively evaluate recommendation harmfulness. Through empirical analysis on these resources, we reveal three alarming findings: harmful content is widespread in recommendation data, users tend to engage with such content, and current methods often facilitate its propagation. To address the identified issue, we propose \baby, a generic and model-agnostic framework that can be seamlessly integrated into existing neural models to mitigate harmful content propagation. We sincerely hope that our benchmark, findings, and framework can inspire future research toward building more responsible, safer recommendation techniques.

\newpage


\bibliographystyle{IEEEtran}

\bibliography{ref}

\appendix

\section{Appendix}
\subsection{Ethics Statement}\label{ethics}
This work aims to raise awareness of the overlooked risk of harmful content propagation in recommender systems and to provide resources and methods for developing harmfulness-aware recommendation technologies. All datasets used in this study are publicly available and already contain harmful content. Our contribution lies solely in the annotation of such content from a harmfulness perspective; we do not create, modify, or amplify any harmful material. 

All data and annotations are intended strictly for scientific research purposes and are prohibited from any commercial use. We fully comply with the data usage policies and terms of service of the original platforms from which the datasets were derived. The content, views, or statements included in the annotated datasets do not reflect the perspectives of the authors or affiliated institutions.

To protect annotators involved in the harmfulness labeling process, particularly when reviewing harmful or distressing content, we implement the following ethical safeguards: (1) obtain informed consent about the nature of the task, including potential exposure to disturbing material; (2) enforce reasonable annotation workloads with weekly limits to manage exposure levels; (3) recommend pausing or discontinuing annotation if discomfort arises; and (4) conduct regular mental wellness check-ins with annotators. These measures are intended to minimize psychological risks and ensure a responsible annotation process.

\subsection{Annotation details}\label{app_annotation}

\subsubsection{Discussion about harmful content definition}\label{app_definition}
Our definition of harmful content is developed by synthesizing existing studies, platform policies, and preliminary analysis. However, harmfulness is inherently context-dependent, and perceptions of what constitutes harmful content may vary across cultures, regions, and time, potentially introducing subjectivity in certain borderline cases. Moreover, harmful content is constantly evolving alongside emerging social trends and new forms of expression, such as novel misinformation topics and evolving slang for offensive language. Therefore, periodic refinement and updates to the definition will be necessary to maintain its relevance and ensure that future benchmarks can adapt to the evolving landscape of harmful content.

\begin{figure*}[t]
  \centering
  \includegraphics[width=0.85\linewidth]{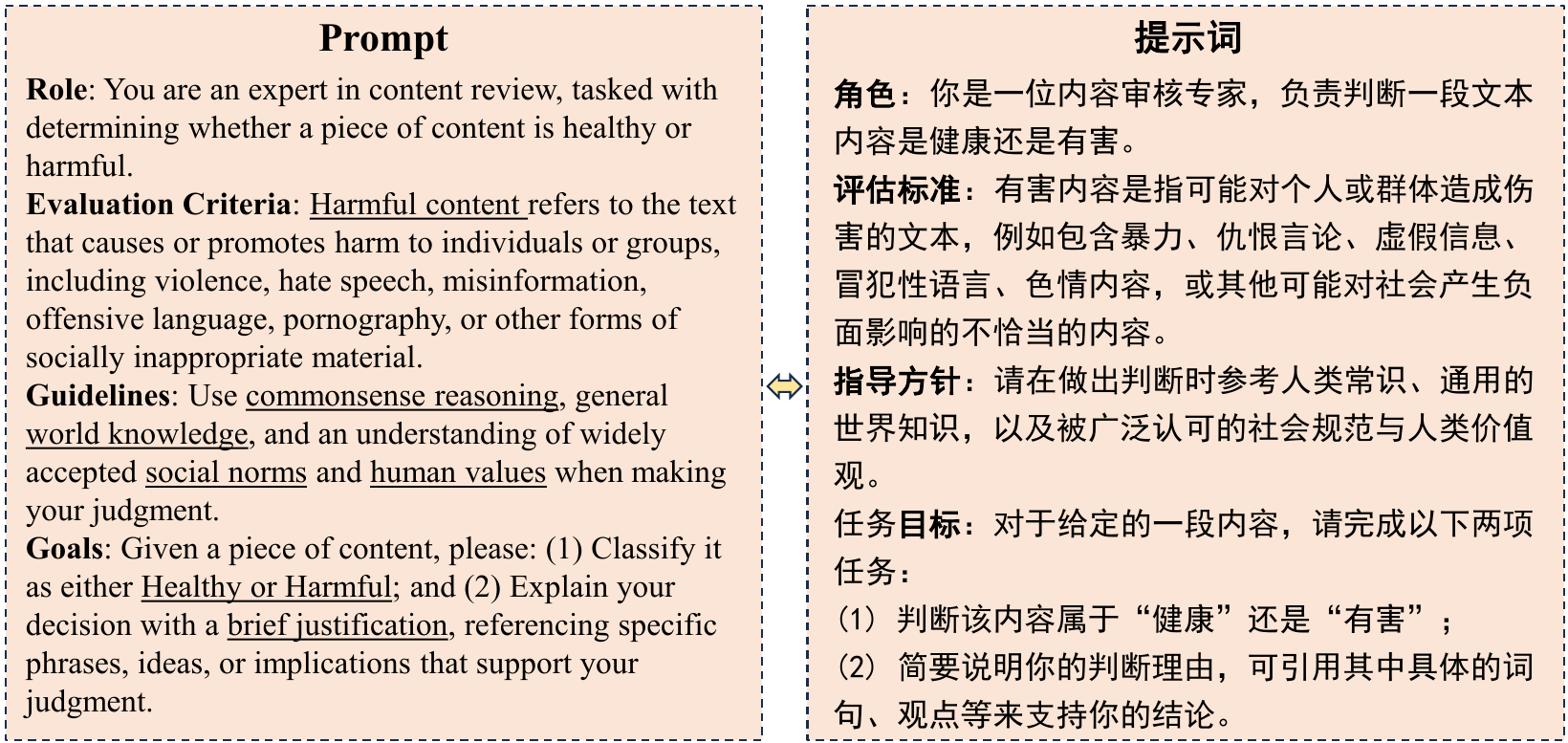}
  \caption{English and Chinese prompts used in the LLM annotation stage for content annotation.}\label{cpropmt}
\end{figure*}

\subsubsection{Data Preparation}\label{app_data}
This work extends three public datasets to examine RS in terms of content harmfulness, including \textbf{TN}, \textbf{QB} and \textbf{EB-NeRD}. These datasets are enriched by annotating harmfulness tags and reasons. TN and QB are two large-scale streaming recommendation datasets containing items such as Chinese news articles, and advertisements~\cite{NineRec}. EB-NeRD~\cite{NeRD} is a News Recommendation Dataset from Ekstra Bladet, used in the RecSys’24 Challenge. It includes a collection of Danish news articles, complete with titles, abstracts, bodies, and metadata like categories. To ensure annotation quality, we utilize translation tools, \ie GPT-4o, to translate Danish content into English for annotation.

\subsubsection{Details about the LLM annotation}\label{app_LLManotation}

In the LLM annotation stage, we leverage the capabilities of large language models to automatically generate initial harmfulness tag and corresponding reason for each item. Given the multilingual nature of our datasets, \ie TN and QB featuring Chinese content, and EB-NeRD comprising English content translated from Danish, we select language-appropriate models to ensure annotation quality. Specifically, we utilize DeepSeek-v3\footnote{\url{https://www.deepseek.com/}} for annotating Chinese content in TN and QB, and GPT-4o\footnote{\url{https://chatgpt.com/}} for processing English content in EB-NeRD. The English prompt of GPT-4o and its Chinese counterpart of DeepSeek-v3 used during the LLM annotation stage are shown in~\fig~\ref{cpropmt}. This modular design enables our LLM-human collaborative annotation paradigm to flexibly incorporate different LLMs based on different requirements, \eg language, thereby enhancing annotation scalability and quality. Importantly, as LLM technology continues to advance, our approach stands to further reduce human annotation workload while maintaining high-quality, diverse, and context-sensitive annotations. In addition, the inputs for EB-NeRD include both the original Danish content and its English translation, as GPT-4o can process multilingual content. The inclusion of the original Danish version contributes to the accurate detection of harmful content.


\subsubsection{Instances of Annotation}\label{app_IO}
\begin{figure*}[ht]
  \centering
  \includegraphics[width=0.85\linewidth]{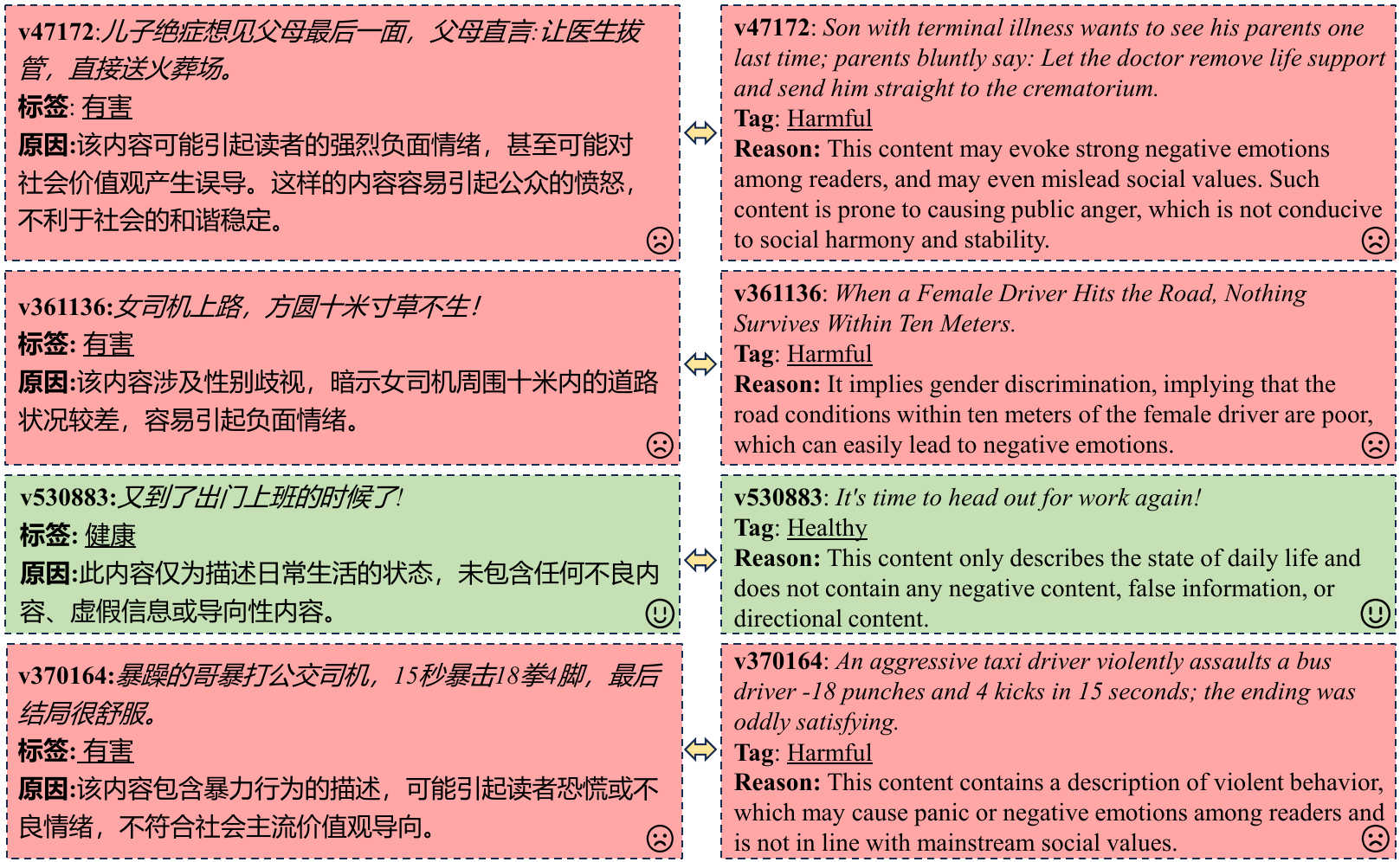}
  \caption{Annotation instances in TN dataset.}\label{IO_TN}
\end{figure*}

\begin{figure*}[ht]
  \centering
  \includegraphics[width=0.85\linewidth]{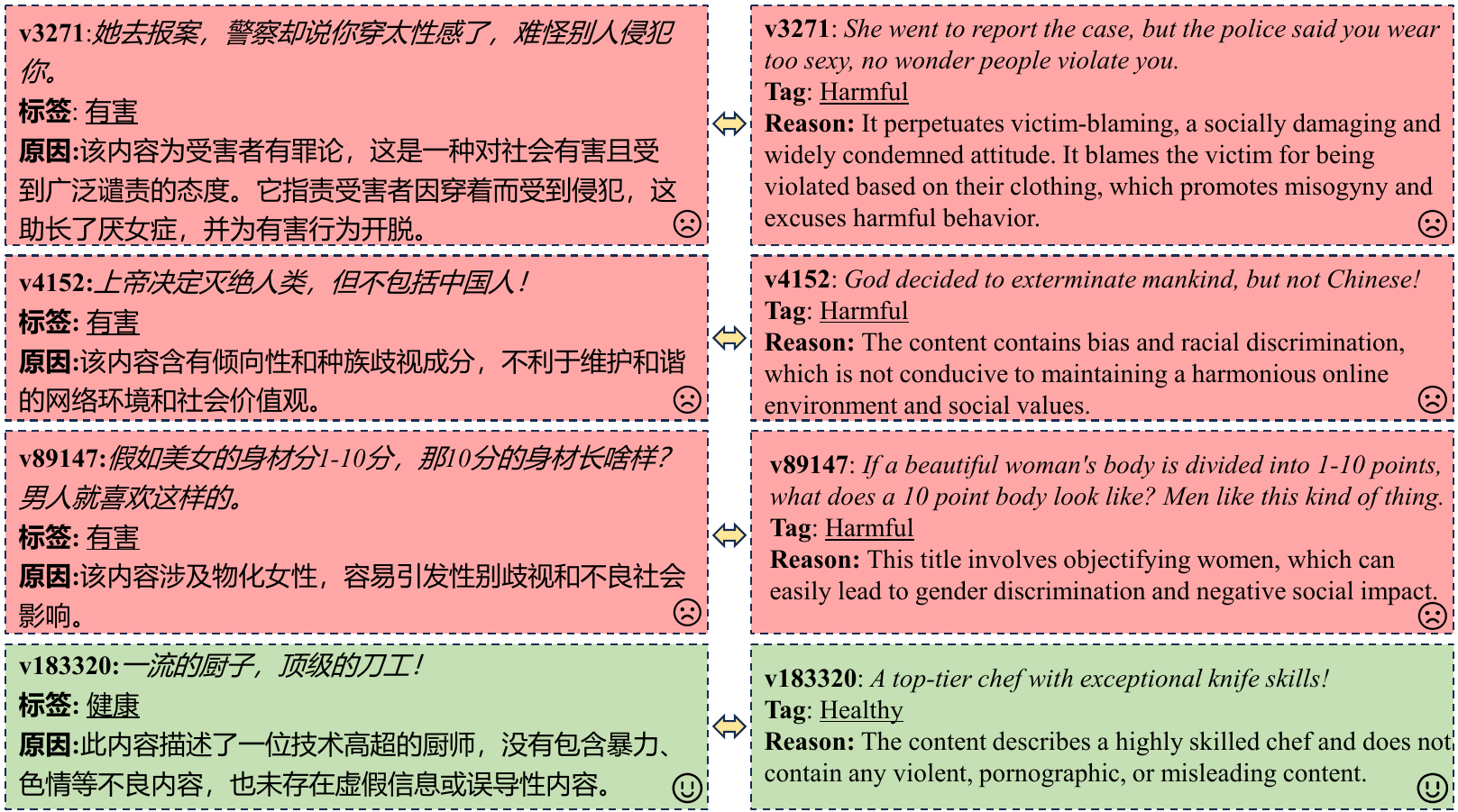}
  \caption{Annotation instances in QB dataset.}\label{IO_QB}
\end{figure*}

\begin{figure*}[ht]
  \centering
  \includegraphics[width=0.85\linewidth]{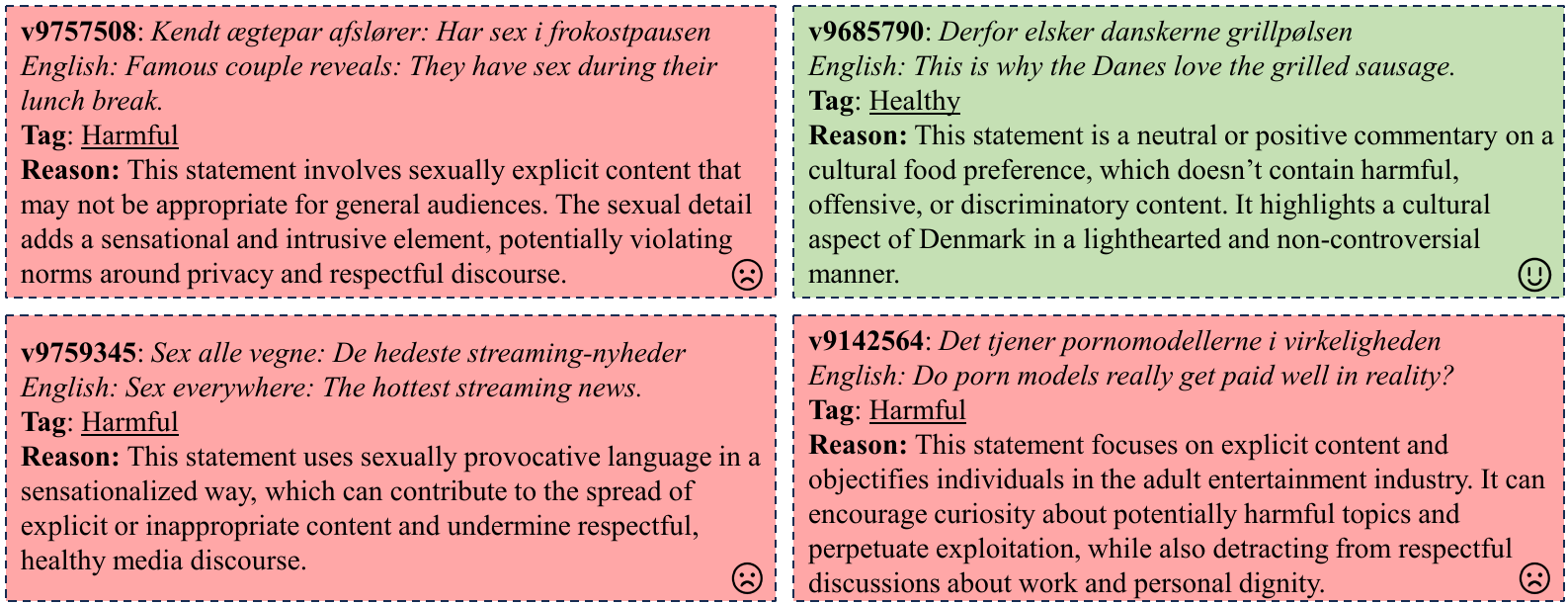}
  \caption{Annotation instances in EB-NeRD dataset.}\label{IO_NERD}
\end{figure*}

To intuitively showcase the effectiveness of our LLM-human collaborative annotation paradigm, we present several instances of TN, QB and EB-NeRD datasets in~\fig~\ref{IO_TN}, \ref{IO_QB} and \ref{IO_NERD}, respectively. Each instance includes the original ID of an item, its content, harmfulness tag, and corresponding reason. In the figure, harmful content is marked with a red background, while healthy content is marked green. These examples highlight the prevalence of harmful content within online platforms, manifesting in forms such as gender discrimination, violence, victim-blaming, racism, objectifying women, incitement to crime, and pornograph. Obviously, exposure to such harmful content, especially at scale, can negatively affect users’ mental well-being, particularly among vulnerable groups like children and adolescents. Furthermore, the widespread propagation of this content can undermine the overall quality and safety of online platforms, posing serious risks to the integrity of the digital ecosystem. Encouragingly, our devised paradigm effectively captures harmfulness-related information. This performance is largely attributed to the LLM’s extensive world knowledge, which enables it to understand relevant cues in content. Besides, our carefully designed prompt provides an informative and structured definition of healthy versus harmful content, effectively activating the LLM’s potential. By requiring LLMs to produce detailed reasons rather than simple binary labels, we enhance both interpretability and reasoning quality, akin to chain-of-thought (CoT)~\cite{Wei@NIPS2022}, addressing the potential hallucination issues of LLMs~\cite{Ji@CS2023}. Moreover, by shifting the bulk of annotation work to LLMs and assigning annotators to focus on quality control, the two-stage paradigm significantly enhances annotation efficiency, demonstrating strong practical potential for extending existing datasets with harmfulness-related information. This strength also positions our work as a solid foundation for building more responsible, harmfulness-aware recommender systems, paving the way for future research and development in this important direction.

\subsubsection{Discussion of Annotation Consistency}\label{app_ac}
To evaluate the reliability of the human annotation process, we compute the Inter-Annotator Agreement (IAA) of the five annotators on the classification of harmfulness tag using Fleiss’ Kappa, which yields a Kappa score of 0.71, indicating a moderate agreement. This level of disagreement is understandable given the subtle and context-dependent nature of harmful content within various online platforms, especially considering that annotators come from diverse regions and hold different cultural backgrounds, which can influence their judgment. Besides, a common source of divergence arises from content that includes humor or sarcasm, which some annotators perceive as healthy while others interpret as harmful. These variations highlight the inherent subjectivity involved in assessing content harmfulness, despite our efforts to standardize annotation guidelines and conduct annotator training sessions. Such findings underscore the complexity of harmful content, highlighting the need for continued investigation into this important topic.

\subsubsection{Error Analysis of LLM Annotation}\label{app_error}
\begin{figure*}[ht]
  \centering
  \includegraphics[width=0.85\linewidth]{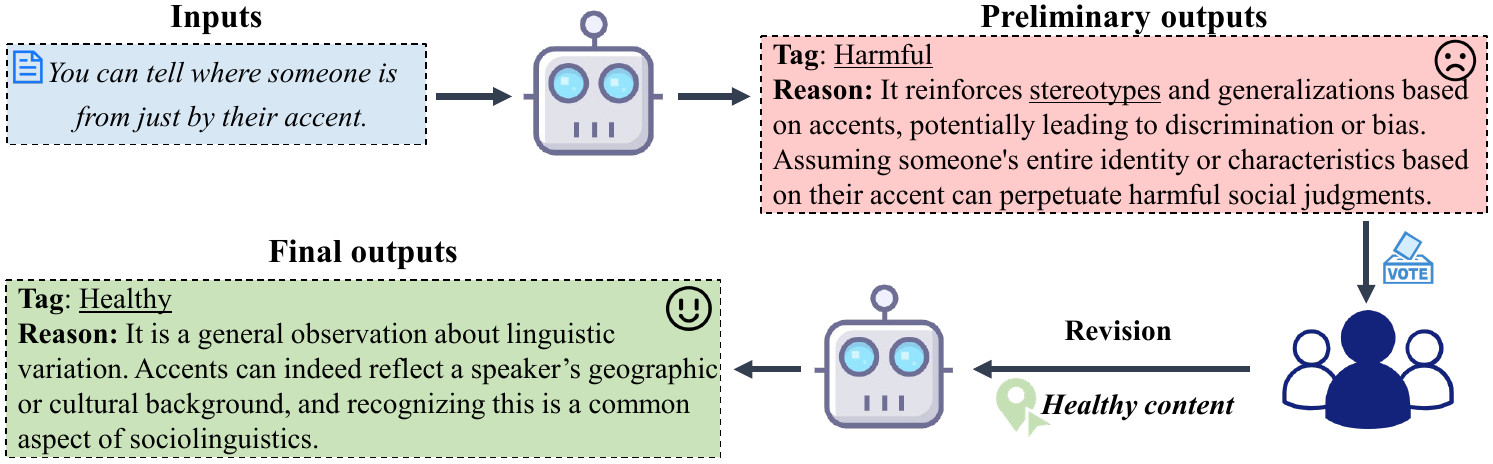}
  \caption{Error case of LLM annotation.}\label{error}
\end{figure*}

An illustrative error case generated in the devised LLM annotation stage is presented in~\fig~\ref{error}. For the input sentence, the LLM initially labeled it as harmful and provided a supporting reason during the LLM annotation stage. However, three out of five human annotators disagreed, determining that the content simply describes a neutral observation about linguistic variation without expressing any hostility or prejudice toward a particular group. Based on the majority judgment, we corrected the label to healthy and re-submitted the content to the LLM, prompting it to generate a revised harmfulness reason in line with the updated classification. This correction was adopted as the final annotation. The LLM’s initial misclassification is understandable, as we explicitly prompt it to adopt a cautious stance to avoid overlooking potential harmful content. This case highlights both the limitations of LLM-only annotation and the strength of our LLM-human collaborative annotation paradigm: while the LLM greatly enhances annotation efficiency, human oversight is essential to ensure annotation accuracy.

\subsection{Discussion about the Proposed Harmfulness Metrics}\label{app_metric}

To quantitatively evaluate the harmfulness of recommendation outputs, we propose two complementary metrics: HP@$k$ and HO@$k$. HP@$k$ calculates the proportion of healthy content among the top-$k$ recommended items, treating all positions in the list equally. In contrast, HO@$k$ emphasizes the position of healthy content by assigning higher weights to items appearing earlier in the ranking, thus capturing the prominence of healthy content exposure. This distinction allows these two metrics to assess recommendation harmfulness from different aspects. For example, consider two recommendation lists: $r^1$ = [+,+,-,-,-] and $r^2$ = [-,-,-,+,+], where ``+'' denotes healthy and ``–'' denotes harmful content. Both lists yield the same HP@5 value of 0.4, indicating that 40\% of their content is healthy. However, in $r^1$, the healthy content appears at the top, making it more visible and impactful to users, whereas in $r^2$, it is buried at the bottom. HO can reflect this difference by rewarding top-ranked healthy items, with HO@5 values of 65.69\% for $r^1$ and 19.71\% for $r^2$. This demonstrates that the proposed HP and HO metrics are complementary. Together, they provide a more comprehensive assessment of the harmfulness of recommendation results. In addition, these two metrics yield the same value when k = 1 or when all items in the recommendation list are healthy or harmful, as position-based weighting in HO becomes irrelevant under these conditions.

\subsection{Details about Representative Recommendation Methods}\label{all_models}

In this work, we select the following 13 representative recommendation models to systematically investigate the harmfulness of recommendation outputs:
\begin{itemize}[leftmargin=*]
    \item GRU4Rec~\cite{GRU4Rec} employs the Gated Recurrent Unit (GRU) to handle the sequential dependencies among items;
    \item NARM~\cite{NARM} utilizes a GRU with an attention mechanism to reveal the main intent of users;
    \item SASRec~\cite{SASRec} leverages a self-attention architecture to model item transitions among user behavior sequences;
    \item BERT4Rec~\cite{BERT4Rec} employs a bidirectional transformer structure to handle the user behavior sequence;
    \item SR-GNN~\cite{SR-GNN} builds item-item graphs based on user behavior sequences and captures complex item relations via GNN;
    \item LESSR~\cite{LESSR} tackles information loss of GNN-based model in the task by introducing shortcut graph attention and edge-order preserving aggregation layers;
    \item PLM-NR~\cite{Wu@SIGIR2021} incorporates pre-trained language models into the content recommendation task;
    \item MSGIFSR~\cite{guo@WSDM2022} divides a user-item interaction sequence into multiple snippets and captures multi-granularity item co-occurrence patterns;
    \item Atten-Mixer~\cite{Zhang@WSDM2023} leverages multi-level user intents to achieve multi-level reasoning on item co-occurrence transitions;
    \item Prompt4NR~\cite{Zhang@SIGIR2023} aims to resolve the content recommendation with prompt learning, where it designs and ensembles a series of prompt templates to make recommendations;
    \item BSARec~\cite{Shin@AAAI2024} addresses two key limitations of Transformer-based models in the task, namely insufficient inductive bias and low-pass filtering properties;
    \item DIMO~\cite{DIMO} enhances user behavior understanding by disentangling ID information and modality-specific signals;
    \item Oracle4Rec~\cite{Xia@WSDM2025} jointly models users' past and future behaviors, aiming to minimize the discrepancy between them to improve prediction accuracy.
    
\end{itemize}

These models span a broad range of architectures and techniques, such as Recurrent Neural Networks (GRU4Rec~\cite{GRU4Rec}, NARM~\cite{NARM}), Attention mechanism (SASRec~\cite{SASRec}, Atten-Mixer~\cite{Zhang@WSDM2023}, BSARec~\cite{Shin@AAAI2024}, DIMO~\cite{DIMO}, Oracle4Rec~\cite{Xia@WSDM2025}), Transformer structures (BERT4Rec~\cite{BERT4Rec}, BSARec~\cite{Shin@AAAI2024}), Graph Neural Networks (SR-GNN~\cite{SR-GNN}, LESSR~\cite{LESSR}, MSGIFSR~\cite{guo@WSDM2022}), as well as large language models (Prompt4NR~\cite{Zhang@SIGIR2023}), and cover major advances in the field from 2016 to now. By evaluating such a diverse set of models, we aim to provide a comprehensive and insightful understanding of how different recommendation techniques perform in terms of the harmfulness perspective. For each model, we process item text using a text encoder (\eg BERT) and utilize the resulting text embeddings to initialize item embeddings. We run each method five times with different random seeds and report the averaged results to ensure the reliability of our evaluation. 

\subsection{Details of Experiment Setup}\label{app_setup}



\subsubsection{Evaluation Metrics}\label{app_setup_evaluation}

\textbf{Accuracy-oriented Metric:}
\textbf{Prec@k}: Precision measures the proportion of cases where the target item appears within the top-$k$ recommendations. It evaluates whether the recommendation list successfully includes the correct item.
\textbf{MRR@k}: Mean Reciprocal Rank highlights the ranking of the target item within the list. It is computed as the mean of the reciprocal ranks of the target item. If the target item’s rank exceeds $k$, its reciprocal rank is set to zero.

\textbf{Harmfulness-oriented Metric:}
This work emphasizes the significance of recommending healthy content in RS. However, existing evaluation metrics do not account for the harmfulness of recommendations. To bridge this gap, we present the HP@$k$ and HO@$k$ metrics, which are detailed in Section \uppercase\expandafter{\romannumeral2}-B.



\begin{figure*}[th]
  \centering
  \includegraphics[width=0.85\linewidth]{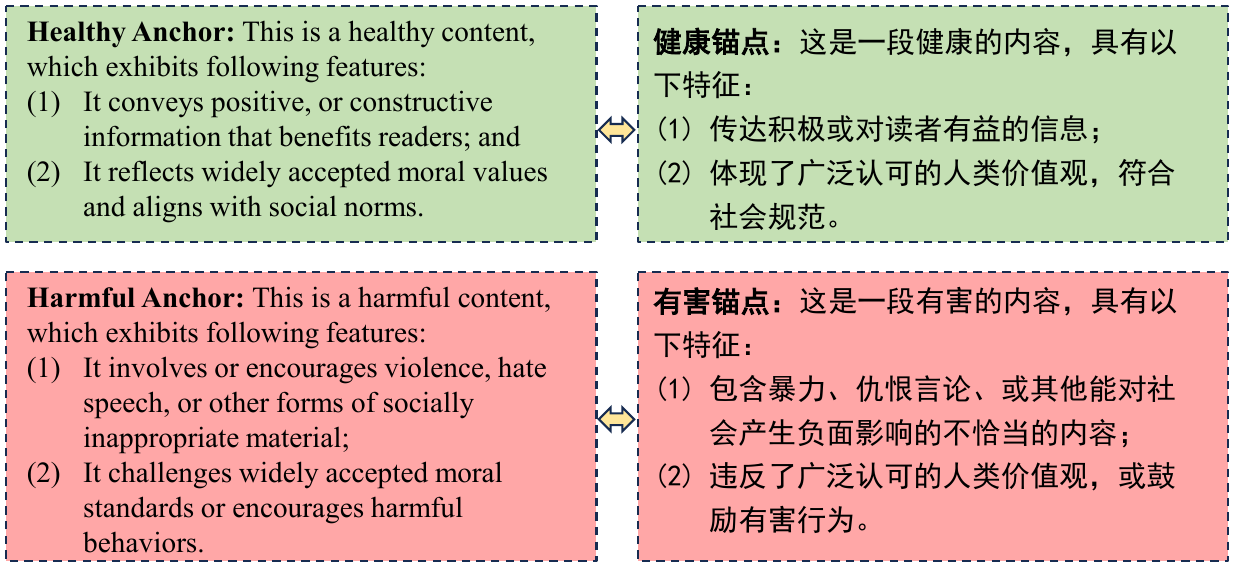}
  \caption{Healthy and harmful anchors in English and Chinese.}\label{canchor}
\end{figure*}

\subsubsection{Base Models}\label{app_setup_baseline}
In this work, we select four base models to evaluate the effect of our proposed framework, \baby, in mitigating the propagation of harmful content, including {GRU4Rec}~\cite{GRU4Rec}, {NARM}~\cite{NARM}, {SASRec}~\cite{SASRec} and {BERT4Rec}~\cite{BERT4Rec}. The details of these models are discussed in Appendix~\ref{all_models}.


These baseline models are widely recognized as benchmarks and foundational architectures for developing recommender systems. Therefore, the effectiveness of \baby on these models suggests its potential applicability to other approaches in the domain. Due to the space limitations, we present the performance of \baby with GRU4Rec, NARM and BERT4Rec in main content of this work, while the results of SASRec is discussed in Appendix~\ref{app_sasrec}.

\begin{table*}[th]
\tabcolsep 0.1in 
  \centering
    \caption{Accuracy and harmfulness performance (\%) of \baby framework on SASRec model. Statistical significance of improvements is determined by the t-test ($p < 0.01$).}
    \begin{tabular}{cc ccc |ccc}  
    \toprule
    
    \multirow{2}{*}{Dataset}    &   \multirow{2}{*}{Method}&  \multicolumn{3}{c}{Accuracy}  & \multicolumn{3}{c}{Harmfulness} \cr
    \cmidrule(lr){3-5}  \cmidrule(lr){6-8}
    & &Prec@1&Prec@5&MRR@5 &HP@1&HP@5&HO@5\cr  
    
    \midrule  
    \multirow{3}*{TN}&
    SASRec &9.10  &\textbf{26.10}  &14.94  &45.58  &46.71   &46.35      \cr
    &\baby &\textbf{10.03}  &25.87  &\textbf{15.10}  &\textbf{62.06}  &\textbf{61.94}   &\textbf{62.58}      \cr
    &$impro.$ &+10.33\%  &-0.88\%  &+1.07\%  &+36.16\%  &+32.61\%   &    +30.52\%\cr
    \midrule  
    \multirow{3}*{QB}&
    SASRec &\textbf{75.47}  &\textbf{80.30}  &\textbf{77.55}  &47.99  &47.12   &47.31      \cr
    &\baby &75.01  &80.13  &77.20  &\textbf{56.52}  &\textbf{66.47}   &\textbf{64.11}      \cr
    &$impro.$ &-0.61\%  &-0.21\%  &-0.45\%  &+17.77\%  &+41.07\%   &+35.51\%      \cr
    \midrule  
    \multirow{3}*{EB-NeRD}&
    SASRec &3.18  &12.21  &6.88  &50.55  &49.84 &50.36     \cr
    &\baby &\textbf{4.83}  &\textbf{14.37}  &\textbf{8.21}  &\textbf{62.37}  &\textbf{60.78}   &\textbf{61.72}     \cr
    &$impro.$ &+52.08\%  &+17.69\%  &+19.33\%  &+23.38\%  &+21.95\%   &+22.56\%      \cr
    
	\bottomrule
    \end{tabular}
    \label{sasrec_results}
\end{table*}

\subsubsection{Details of the Harmful and Healthy Anchor Construction}
In \baby, we design two types of content anchors, \ie harmful and healthy anchors, to enhance its ability to identify content harmfulness. The harmful anchor serves as a clean and explicitly harmful signal, while the healthy anchor provides a clean and fully healthy signal. To construct the harmful anchor, we directly use the harmful semantics specified in our harmful content definition. For the healthy anchor, we instruct an LLM to generate a healthy counterpart based on the constructed harmful anchor. Specifically, the instruction is as follows:

\textbf{Role}: You are an expert in content generation. Please generate a piece of healthy content that serves as a clean and unambiguous healthy reference signal. 

\textbf{Guidelines}: The generated content should be a healthy counterpart to the following harmful anchor: [Harmful Anchor].

\textbf{Requirement}: Generate natural, coherent, and brief content that avoids all harmful characteristics and promotes positive, constructive, and socially responsible values.

Here, the harmful anchor is provided as input to guide the generation of its healthy counterpart. The resulting healthy content is then used as the healthy anchor. The healthy and harmful anchors used for English and Chinese content are illustrated in~\fig~\ref{canchor}.

\subsubsection{Implementation Details}\label{app_setup_implementation}

The hardware used in this study is an NVIDIA GeForce RTX 4090 GPU with 24GB of memory, and the software environment is based on Python 3.9.5 and PyTorch 1.12.0. To ensure a fair comparison across all methods, we fix the embedding dimension at $d=100$ and set the mini-batch size to 512. Model training is conducted using the Adam optimizer with a learning rate of $1 \times 10^{-3}$. For all baseline methods, hyper-parameters are tuned via grid search based on their performance in terms of Prec@5 on the validation set. To prevent overfitting, we employ early stopping with a patience of 10 epochs according to models' performance on Prec@5. The base models and their enhanced versions with \baby share identical experimental settings. The balancing coefficient $\lambda$ in \baby is searched from \{0.001, 0.01, 0.1, 1\}. The pre-trained text encoder $textEncoder(*)$ is instantiated with BERT~\cite{BERT}, and dot-product is used as the similarity function $<*>$. We run each method 5 times independently, and use the average performance as the final results.
All resources of this work are available online\footnote{\url{https://github.com/Zhang-xiaokun/MitiHarm}}.

\subsection{Instantiation of \baby on Additional Models}\label{app_sasrec}

We initially instantiated our \baby framework using three representative models: GRU4Rec, NARM, and BERT4Rec. To further verify the generalizability of \baby, we integrate it into another popular model, \ie SASRec, which is based on the self-attention mechanism. The performance comparison between SASRec and its \baby-enhanced version under accuracy and harmfulness perspectives is reported in Table~\ref{sasrec_results}. The results demonstrate that \baby significantly mitigates the harmfulness of SASRec, \eg achieving up to 41.07\% and 35.51\% improvements in terms of HP@5 and HO@5 on QB dataset, respectively. This confirms the effectiveness of \baby in mitigating the risk of harmful content propagation in current recommender systems. Moreover, \baby also achieves impressive performance in prediction accuracy. Remarkably, in certain scenarios, it even outperforms SASRec by a large margin. For example, on EB-NeRD dataset, the accuracy improvement exceeds 50\% in terms of Prec@1. These findings underscore the strong robustness and applicability of \baby, showcasing its ability to mitigate recommendation harmfulness while preserving or even improving recommendation relevance.

\subsection{More Discussion}
In this work, we reveal that existing recommender systems tend to expose users to harmful content. Given the increasingly important role of RSs in modern information distribution, such recommendation behavior can facilitate the propagation of harmful content. Nevertheless, information propagation in real-world environments is a complex process that is inherently dynamic and cascade-level, and can be influenced by various factors, such as user preferences and social connections, content characteristics, user-generated responses, and interactions across multiple recommendation rounds. Investigating these dynamics would require longitudinal interaction data and/or controlled counterfactual experiments that track users and content over multiple rounds of exposure and interaction. In contrast, our work focuses specifically on harmful-content propagation from the recommendation-exposure perspective: by reducing the amount and ranking prominence of harmful content presented by recommender systems, our approach reduces an important pathway through which harmful content reaches users and may subsequently propagate. We leave the investigation of dynamic, cascade-level propagation effects and their causal relationships to future work.



\end{document}